\documentclass[showpacs,preprintnumbers,amsmath,amssymb]{revtex4}

\usepackage{epsfig}
\usepackage{graphicx}
\usepackage{dcolumn}
\usepackage{bm}

\begin{document}


\title{Wave-Function Factorization of the Normal-Ordered 1D Hubbard Model
for Finite Values
 of the On-site Repulsion $U$}
\author{J. M. P. Carmelo}
\affiliation{Department of Physics, Massachusetts Institute of Technology,
Cambridge, Massachusetts 02139-4307}
\affiliation{GCEP-Center of Physics, University of Minho, Campus
Gualtar, P-4710-057 Braga, Portugal}
\date{23 May 2003}


\begin{abstract}
In this paper we find that in the thermodynamic limit and for the
the ground-state normal-ordered 1D Hubbard model the wave function
of the excited energy eigenstates which span the Hilbert subspace
where the finite-number-electron excitations are contained factorizes
for all values of the on-site Coulombian repulsion $U$. This factorization
results from the absence of residual energy interactions for the
pseudofermions whose occupancy configurations describe these
states. Our study includes the introduction of the pseudoparticle -
pseudofermion unitary transformation and of an operator algebra
for both the pseudoparticles and the pseudofermions. As the
corresponding pseudoparticles, the $c\nu$ pseudofermions (and
$s\nu$ pseudofermions) are $\eta$-spin zero $2\nu$-holon composite
quantum objects (and spin zero $2\nu$-spinon composite quantum
objects) where $\nu=1,\,2,...$. The pseudofermion description is
the most suitable for the study of the finite-energy spectral
properties of the model.
\end{abstract}


\maketitle
\section{INTRODUCTION}

Recently there has been a renewed experimental interest in the
exotic one-electron and two-electron spectral properties of
quasi-1D materials \cite{spectral0,Menzel,Hasan,Ralph}. Some of
these experimental studies observed unusual
finite-energy/frequency spectral properties, which are far from
being well understood. For low values of the energy, the
microscopic electronic properties of these materials are usually
described by systems of coupled chains. For finite values of the
energy larger than the transfer integrals for electron hopping
between the chains, the one-dimensional (1D) Hubbard model is
expected to provide a good description of the physics of these
materials \cite{Menzel,Hasan,Ralph}. This is confirmed by the
recent quantitative studies of Refs.
\cite{spectral0,spectral,super}. Similar unusual spectral
properties observed in two-dimensional (2D) high-$T_c$
superconductors could result from effective quasi-1D charge and
spin transport \cite{Menzel,Granath,Zaanen}. The 1D Hubbard model
is also suitable for the description of the microscopic
mechanisms behind the spectral properties of the new quantum
systems described by cold fermionic atoms on an optical lattice.
Indeed, following the experimental studies of strongly correlated
quantum systems of ultra cold bosonic atoms held in optical
lattices \cite{Greiner}, new experiments involving cold
fermionic atoms (such as $^6Li$) on a 1D optical lattice formed by
interfering laser fields are in progress. Provided that the electrons
are replaced by such atoms, that system can be described by the above model.
However, the non-perturbative nature of the 1D Hubbard model implies that the
electronic creation and annihilation operators do not provide a
suitable operational description for the study of the
finite-energy spectral properties. Thus, the first step for the
study of these properties is the introduction of a suitable
operational description. Except in the limit of infinite on-site
Coulombian repulsion $U\rightarrow\infty$
\cite{Ogata,Ricardo,Penc95,Penc96}, the introduction of such a
description is an open problem of great physical interest. For low
values of energy, useful information about the effects of the
non-perturbative electronic correlations is provided by
two-component conformal-field theory
\cite{Belavin,Frahm,Carmelo91,Carmelo92,Carmelo97pp}.
Unfortunately, that method does not apply for finite values of
energy.

In view of the above-mentioned finite-energy/frequency
spectral properties observed in real experiments, which are far
from being well understood, efforts towards the introduction of a
suitable operational description to deal with the finite-energy
problem are welcome. In this paper we introduce an operational
representation for the 1D Hubbard model
\cite{Lieb,Takahashi,Martins98,Hubbard} in terms of the {\it pseudofermions}.
In contrast to the related pseudoparticles of Refs. \cite{I,II}, the
pseudofermions have no residual energy interactions. We find that in the
thermodynamic limit the wave function of the excited energy eigenstates
which span the Hilbert subspace where the finite-number-electron excitations
are contained factorizes into separated contributions corresponding to
different pseudofermion branches. (A finite-number-electron excitation is
generated by application onto the ground state of operators whose
expression involves the product of a finite number of electronic creation
and/or annihilation operators.) That factorization occurs for all
values of the on-site Coulombian repulsion $U$ of the ground-state
normal-ordered 1D Hubbard model. The pseudofermion operational
description is closely related to the pseudoparticle
representation \cite{I,II}, and is the natural starting point for studies of the
finite-energy/frequency finite-number-electron spectral properties.
Specifically, our starting point is a holon, spinon, and $c0$
pseudoparticle representation recently introduced in Ref.
\cite{I}, which refers to the whole Hilbert space of the model.
The relation between the original electrons and these elementary
quantum objects involves the concept of {\it rotated electron}.
The rotated electrons are related to the electrons by a unitary
transformation first introduced in Ref. \cite{Harris}. The
concepts of local pseudoparticle and effective lattice widely used
in this paper are introduced in Ref. \cite{IIIb}. As a result of
the wave-function factorization, the pseudofermion description is
more suitable for the study of the overlap between finite-number-electron
excitations and the energy eigenstates than the pseudoparticle
representation.

As further discussed in Sec. V, the pseudofermion operational
description introduced in this paper and the above-mentioned
associated wave-function factorization are used elsewhere in the
construction of a {\it pseudofermion dynamical theory} \cite{V,V-1}.
Preliminary applications of such a theory led to a successful
description of the finite-energy spectral properties observed in
low-dimensional complex materials
\cite{spectral0,spectral,super}. In spite of the absence of
residual energy interactions, the pseudofermions are shown to
be active scatterers and scattering centers in Ref. \cite{S-P},
where the corresponding pseudofermion scattering theory is
introduced.

The paper is organized as follows: In Sec. II the 1D Hubbard model
and the rotated electrons are introduced and useful information
about the pseudoparticle representation is provided. This includes
the introduction of the operator algebra for the pseudoparticles.
The pseudofermion description and the relationship between
pseudoparticle and pseudofermion operators are introduced and
discussed in Sec. III, as well as the pseudofermion anticommuting
algebra. In Sec. IV the pseudofermion energy and momentum
spectra are studied and the factorization of the finite-number-electron
Hilbert subspace of the ground-state normal-ordered
1D Hubbard model is introduced and discussed. Finally, in Sec. V the
discussion and the concluding remarks are presented.

The pseudofermion operational description and all other issues and
concepts associated with that description introduced and studied
in Secs. III and IV correspond to a novel approach to the present
quantum problem. On the other hand, the information about the
holon, spinon, and pseudoparticle representations presented in
Sec. II follows the results of Ref. \cite{I} and is a review
and/or different interpretation of known
methods and concepts that are needed for the introduction of the
pseudofermion description.

\section{THE 1D HUBBARD MODEL, ROTATED ELECTRONS, AND THE PSEUDOPARTICLE
DESCRIPTION}

\subsection{THE 1D HUBBARD MODEL AND ROTATED ELECTRONS}

In a chemical potential $\mu $ and magnetic field $H$ the 1D
Hubbard Hamiltonian can be written as,

\begin{equation}
\hat{H}={\hat{H}}_{SO(4)} + \sum_{\alpha =c,\,s}\mu_{\alpha}\,
{\hat{S}}_{\alpha}^z \, , \label{H}
\end{equation}
where the Hamiltonian ${\hat{H}}_{SO(4)} = {\hat{H}}_{H} -
(U/2)\,\hat{N} + (U/4)\,N_a$ has $SO(4)$ symmetry and
${\hat{H}}_{H} = \hat{T}+U\,\hat{D}$ is the ``simple" Hubbard
model. Here $\hat{T}=-t\sum_{\sigma=\uparrow ,\,\downarrow
}\sum_{j=1}^{N_a}\Bigl[c_{j,\,\sigma}^{\dag}\,c_{j+1,\,\sigma} +
h. c.\Bigr]$ is the {\it kinetic-energy} operator and $\hat{D} =
\sum_{j=1}^{N_a}c_{j,\,\uparrow}^{\dag}\,c_{j,\,\uparrow}\,
c_{j,\,\downarrow}^{\dag}\,c_{j,\,\downarrow} =
\sum_{j=1}^{N_a}\hat{n}_{j,\,\uparrow}\,\hat{n}_{j,\,\downarrow}$
is the electron double-occupation operator. On the right-hand side
of Eq. (\ref{H}), $\mu_c=2\mu$, $\mu_s=2\mu_0 H$, $\mu_0$ is the
Bohr magneton, and the number operators, ${\hat{S }}_c^z=-{1\over
2}[N_a-\hat{N}]$ and  ${\hat{S }}_s^z= -{1\over
2}[{\hat{N}}_{\uparrow}- {\hat{N}}_{\downarrow}]$ are the diagonal
generators of the $\eta$-spin and spin $SU(2)$ algebras
\cite{HL,Yang89,Essler}, respectively. The Hamiltonian
$\hat{H}_{SO(4)}$ of Eq. (\ref{H}) commutes with the six
generators of the $\eta$-spin and spin $SU(2)$ algebras and has
$SO(4)$ symmetry \cite{HL,Yang89,Essler}. (The off-diagonal
generators of these two $SU(2)$ algebras are given in Eqs. (7) and
(8) of Ref. \cite{I}, respectively.) The number of lattice sites $N_a$
is considered to be large. The above electronic number
operators read ${\hat{N}}=\sum_{\sigma=\uparrow ,\,\downarrow
}\,\hat{N}_{\sigma}$ and ${\hat{N}}_{\sigma}=\sum_{j=1}^{N_a}
\hat{N}_{j,\,\sigma}$, where $\hat{N}_{j,\,\sigma}
= c_{j,\,\sigma }^{\dagger}\,c_{j,\,\sigma }$ counts
the number of spin-projection $\sigma$ electrons
at real-space lattice site $j$. Moreover, the operator
$c_{j,\,\sigma}^{\dagger}$ (and $c_{j,\,\sigma}$) creates (and
annihilates) a spin $\sigma $ electron at lattice site
$j=1,2,...,N_a$. The electronic lattice constant is denoted by $a$ and
the lattice length by $L=N_a\,a$ and periodic boundary
conditions and the thermodynamic limit $L\rightarrow\infty$
are assumed.

The momentum operator is given by $\hat{P} = \sum_{\sigma=\uparrow
,\,\downarrow }\sum_{k}\, \hat{N}_{\sigma} (k)\, k$ where the
spin-projection $\sigma$ momentum distribution operator reads
$\hat{N}_{\sigma} (k) = c_{k,\,\sigma }^{\dagger }\,c_{k,\,\sigma }$
and the operator $c_{k,\,\sigma}^{\dagger}$ (and $c_{k,\,\sigma}$) creates
(and annihilates) a spin-projection $\sigma $ electron of momentum $k$. The
operators $c_{k,\,\sigma}^{\dagger}$ and $c_{k,\,\sigma}$ are
related to the above operators $c_{j,\,\sigma}^{\dagger}$ and
$c_{j,\,\sigma}$ by the Fourier transforms
$c_{k,\,\sigma}^{\dagger} =
[1/\sqrt{L}]\sum_{j=1}^{N_a}e^{ik\,aj}\, c_{j,\,\sigma}^{\dagger}$
and $c_{k,\,\sigma} = [1/\sqrt{L}]\sum_{j=1}^{N_a}e^{-ik\,aj}\,
c_{j,\,\sigma}$, respectively.

Throughout this paper units of Planck constant one are used and
the electronic charge is denoted by $-e$. The Bethe-ansatz solvability
of the 1D Hubbard model (\ref{H}) is restricted to the Hilbert
subspace spanned by the lowest-weight states (LWSs)
\cite{Lieb,Takahashi} or highest-weight states (HWSs)
\cite{Martins98} of the $\eta$-spin and spin algebras, that is by
the states whose $S_{\alpha}$ and $S_{\alpha}^z$ numbers are such
that $S_{\alpha}= -S_{\alpha}^z$ or $S_{\alpha}=S_{\alpha}^z$,
respectively, where $\alpha =c$ for $\eta$-spin and $\alpha =s$
for spin. In this paper the $\eta$-spin and spin LWSs
description of the Bethe-ansatz solution is used. In this case, that
solution describes energy eigenstates with electronic densities
$n=N/L$ and spin densities $m=[N_{\uparrow}-N_{\downarrow}]/L$ in
the domains $0\leq na \leq 1$ and $0\leq ma \leq na$, respectively.
Some of the results presented in the paper correspond to the ranges
$0< na < 1$ and $0<ma< na$. The description of the states corresponding
to the extended domains $0\leq na \leq 1$\, ; $1\leq na \leq 2$ and
$-na\leq ma \leq na$\, ; $-(2-na)\leq ma \leq (2-na)$, respectively,
is achieved by application onto the latter states of off-diagonal
generators of the $\eta$-spin and spin $SU(2)$ algebras
\cite{I,Essler}.

The electron - rotated-electron unitary transformation maps the
electrons onto rotated electrons such that rotated-electron double
occupation, no occupation, and spin-up and spin-down single
occupation are good quantum numbers for all values of $U/t$.
The lattice occupied by rotated electrons is identical
to the original electronic lattice. The electrons that occur in
the 1D Hubbard model (\ref{H}) are called $c_{j,\,\sigma}^{\dag}$,
while the operator ${\tilde{c}}_{j,\,\sigma}^{\dag}$ such that
${\tilde{c}}_{j,\,\sigma}^{\dag} =
{\hat{V}}^{\dag}(U/t)\,c_{j,\,\sigma}^{\dag}\,{\hat{V}}(U/t)$
represents the rotated electrons, where ${\hat{V}}(U/t)$ denotes
the electron - rotated-electron unitary operator. Similarly,
$c_{j,\,\sigma}^{\dag} =
{\hat{V}}(U/t)\,{\tilde{c}}_{j,\,\sigma}^{\dag}\,{\hat{V}}^{\dag}(U/t)$.
Note that $c_{j,\,\sigma}^{\dag}$ and
${\tilde{c}}_{j,\,\sigma}^{\dag}$ are only identical in the
$U/t\rightarrow\infty$ limit where electron double occupation
becomes a good quantum number. The unitary operators
${\hat{V}}^{\dag}(U/t)$ and ${\hat{V}}(U/t)$ are uniquely defined
for all values of $U/t$ by Eqs. (21)-(23) of Ref. \cite{I}. The
electron - rotated-electron unitary transformation was introduced
in Ref. \cite{Harris}. The rotated-electron double occupation
operator $\tilde{D}$ given in Eq. (20) of Ref. \cite{I} commutes
with the 1D Hubbard model. Thus, the rotated-electron double
occupation $D_r$ is a good quantum number for all values of $U/t$.

\subsection{THE PSEUDOPARTICLE OPERATORS}

According to the studies of Ref. \cite{I}, there is an infinite
number of pseudoparticle branches: the $c0$ pseudoparticles and
the $\alpha\nu$ pseudoparticles such that $\alpha =c,\,s$ and
$\nu=1,2,...$. (Here we call $c0$ pseudoparticles, the $c$
pseudoparticles of Ref. \cite{I}. Thus, within our notation, the
general designation of $\alpha\nu$ pseudoparticle refers to the
$\alpha\nu =c\nu$ branches such that $\nu =0,1,2,...$ and $\alpha\nu =s\nu$
branches such that $\nu =1,2,...$.) It is shown in Ref. \cite{I}
that for $\nu >0$ the $c\nu$ pseudoparticles and $s\nu$
pseudoparticles are $2\nu$-holon and $2\nu$-spinon composite
objects, respectively. Throughout this paper we follow the
notation of that reference and denote the holons and spinons
according to their value of $\eta$-spin projection $\pm 1/2$ and
spin projection $\pm 1/2$, respectively. The $\pm 1/2$ holons
(and $\pm 1/2$ spinons)
which are not part of $2\nu$-holon composite $c\nu$
pseudoparticles (and $2\nu$-spinon composite $s\nu$
pseudoparticles) are called $\pm 1/2$ Yang holons (and $\pm 1/2$
HL spinons). In the designations {\it HL spinon} and {\it Yang
holon}, HL stands for Heilmann and Lieb and Yang refers to C. N.
Yang, respectively, who are the authors of Refs. \cite{HL,Yang89}.

The construction of the pseudofermion operational description
requires the use of an operator representation for the $\alpha\nu$
pseudoparticles, which is introduced in this section. An
operational description for the pseudoparticles of bare-momentum
$q$ was introduced in Ref. \cite{Carmelo97}. However, that
description did not take into account the holon (and spinon)
composite character of the $c\nu$ pseudoparticles (and $s\nu$
pseudoparticles). Another limitation was the lack of a
representation for the pseudoparticle operators in terms of
spatial coordinates. The concepts of local $\alpha\nu$
pseudoparticle and effective $\alpha\nu$ lattice are summarized below.

Generation and removal of pseudoparticles is in general associated
with creation and/or annihilation of electrons. Yet there are also
transitions which change the numbers of these quantum objects at
constant spin-projection $\sigma$ electron numbers. Let us introduce the
bare-momentum $\alpha\nu$ pseudoparticle creation (and
annihilation) operator $b^{\dag }_{q,\,\alpha\nu}$ (and
$b_{q,\,\alpha\nu}$) which creates (and annihilates) a $\alpha\nu$
pseudoparticle of bare momentum $q$. In addition, the
local $\alpha\nu$ pseudoparticle creation operator $b^{\dag
}_{x_j,\,\alpha\nu}$ and annihilation operator
$b_{x_j,\,\alpha\nu}$ are also introduced. These bare-momentum
and local pseudoparticle operators are related as follows,

\begin{equation}
b^{\dag }_{q,\,\alpha\nu} =
{1\over\sqrt{L}}\sum_{j=1}^{N^*_{\alpha\nu}}e^{iq\,x_j}\, b^{\dag
}_{x_j,\,\alpha\nu} \, ; \hspace{1cm} b_{q,\,\alpha\nu} =
{1\over\sqrt{L}}\sum_{j=1}^{N^*_{\alpha\nu}}e^{-iq\,x_j}\,
b_{x_j,\,\alpha\nu} \, . \label{elop}
\end{equation}
The local $\alpha\nu$ pseudoparticle creation (and annihilation)
operator $b^{\dag }_{x_j,\,\alpha\nu}$ (and $b_{x_j,\,\alpha\nu}$)
creates (and annihilates) a local $\alpha\nu$ pseudoparticle at
the effective $\alpha\nu$ lattice site of spatial coordinate $x_j
=a_{\alpha\nu}\,j$, where $j=1,2,...,N^*_{\alpha\nu}$,
$N^*_{\alpha\nu}$ is the number of sites defined in Eqs. (B6)-(B8)
and (B11) and of Ref. \cite{I}, and

\begin{equation}
a_{\alpha\nu} = a\,{N_a\over N^*_{\alpha\nu}} =  {L\over
N^*_{\alpha\nu}} \, , \label{aan}
\end{equation}
is the effective $\alpha\nu$ lattice constant. There
is an effective $\alpha\nu$ pseudoparticle lattice for each
$\alpha\nu$ pseudoparticle branch \cite{IIIb}. This lattice has
the same length $L=N^*_{\alpha\nu}\,a_{\alpha\nu}$ as the original
real-space lattice. For the $\alpha\nu$ pseudoparticle bands such
that $\nu>0$, the discrete bare-momentum values $q_j$ are
distributed symmetrically relative to zero, and are such that
$\vert q_j \vert \leq q_{\alpha\nu}$. The two bare momentum values
$q=\pm q_{\alpha\nu}$ limit the $\alpha\nu$ pseudoparticle {\it Brillouin
zone}, where $q_{\alpha\nu} = [\pi/
a_{\alpha\nu}][1-1/N^*_{\alpha\nu}]$. It follows from the
expressions given in Eq. (5) of Ref. \cite{spectral} for the
ground-state values of the number $N^*_{\alpha\nu}$ that for the
ground state the effective $\alpha\nu$ lattice constants
(\ref{aan}) are given by \cite{IIIb},

\begin{equation}
a_{c0}^0 = a \, ; \hspace{1cm} a_{c\nu}^0 = {1\over \delta} \, ;
\hspace{1cm} a_{s1}^0 = {1\over n_{\uparrow}} \, ; \hspace{1cm}
a_{s\nu}^0 = {1\over m} \, , \label{acanGS}
\end{equation}
where $\delta =(1/a-n)$ is the doping concentration. The meaning
of the divergences in the value of the constants
$a_{\alpha\nu}^0$ defined in Eq. (\ref{acanGS}) is that the
corresponding effective $\alpha\nu$ lattice has no sites, {\it
i.e.} $N^{0,*}_{\alpha\nu}=0$ and, therefore, does not exist for
the ground state. This is the case of the effective $c\nu$
lattices for half filling when $\nu>0$ and of the effective $s\nu$
lattices for zero spin density when $\nu>1$. It follows that such
singularities just indicate the collapse of the corresponding
effective $\alpha\nu$ lattice. This is one of the reasons why some
of the expressions given in this paper refer to electronic densities
$0<na<1$ and spin densities $0<ma<na$, such that all ground-state effective
$\alpha\nu$ lattice constants (\ref{acanGS}) have finite values.

As found below, the effective $\alpha\nu$ pseudoparticle and
$\alpha\nu$ pseudofermion lattices are identical. From use of
expressions (\ref{acanGS}) for the ground-state effective-lattice
constants $a_{\alpha\nu}^0$, one can write the ground-state number
$N^{0,*}_{\alpha\nu}$ given in Eq. (5) of Ref. \cite{spectral} as
$N^{0,*}_{\alpha\nu}=L/a_{\alpha\nu}^0$. Except for terms of order
$1/L$, the limiting bare-momentum values defined in Eqs. (B14),
(B16), and (B17) of Ref. \cite{I} simplify and are given by
$q^0_{\alpha\nu} = \pi/a_{\alpha\nu}^0$. The ground-state
expressions for these $\alpha\nu$ pseudoparticle limiting
bare-momentum values can be rewritten in terms of the original
electronic lattice constant $a$ as,

\begin{equation}
q^0_{c0} = \pi/a \, ; \hspace{0.5cm} q^0_{s1} = k_{F\uparrow} \, ;
\hspace{0.5cm} q^0_{c\nu} = [\pi/a -2k_F] \, , \hspace{0.3cm} \nu
>0 \, ; \hspace{0.5cm} q^0_{s\nu} =
[k_{F\uparrow}-k_{F\downarrow}] \, , \hspace{0.3cm} \nu >1 \, .
\label{qcanGS}
\end{equation}

The conjugate variable of the bare-momentum $q_j$ of the
$\alpha\nu$ pseudoparticle branch is the above space coordinate
$x_j$ of the corresponding effective $\alpha\nu$ lattice. This is
different to the electronic operators, where the conjugate
variable of the momentum $k_j$ is the space variable of the
original electronic lattice. In reference \cite{IIIb}, the
pseudoparticle site distribution configurations in the effective
$\alpha\nu$ lattices that describe the energy eigenstates are
related to the corresponding rotated-electron site distribution
configurations.

The $\alpha\nu$ pseudoparticle bare-momentum distribution
functions $N_{\alpha\nu}(q)$ play an important role in the
pseudoparticle description \cite{I,II}. These functions are for
all energy eigenstates the eigenvalues of the following
pseudoparticle bare-momentum distribution operators,

\begin{equation}
\hat{N}_{\alpha\nu}(q)= b^{\dag
}_{q,\,\alpha\nu}\,b_{q,\,\alpha\nu} \, . \label{Nanop}
\end{equation}
The pseudoparticles obey a Pauli principle relative to the
bare-momentum occupancy configurations, {\it i.e.} a discrete
bare-momentum value $q_j$ can either be unoccupied or singly
occupied by a pseudoparticle. Thus, the bare-momentum distribution
functions read $N_{\alpha\nu} (q_j)= 1$ for occupied discrete
bare-momentum values $q_j$ and $N_{\alpha\nu} (q_j)= 0$ for
unoccupied discrete bare-momentum values $q_j$. Each LWS is
uniquely specified by the values of the set of distribution
functions $\{N_{\alpha\nu}(q)\}$ such that $\nu =0,1,2,...$ for
$\alpha =c$ and $\nu =1,2,...$ for $\alpha =s$. Physical
quantities such as the energy, depend on the values of these
distribution functions and numbers through the rapidity momentum
functional $k(q)$ and rapidity functionals $\Lambda_{c\nu}(q)$ and
$\Lambda_{s\nu}(q)$. The value of these functionals is uniquely
provided by solution of the functional integral equations
(13)-(16) of Ref. \cite{I}. In these equations $k_{c\nu}(q)$ is
the $c\nu$ rapidity-momentum functional and the limiting
bare-momentum values $q_{c0}^{\pm}$ and $q_{\alpha\nu}$ where
$\alpha =c,\,s$ and $\nu=1,2,...$ are given in Eqs. (B14), (B16),
and (B17) of the same reference. [For the particular case of the
ground state these values are provided in Eq. (\ref{qcanGS}).] The
above integral equations correspond to a functional representation
of the thermodynamic Bethe-ansatz equations introduced by
Takahashi \cite{Takahashi}. The rapidity-momentum functional is
real and the rapidity functionals are the real part of Takahashi's
ideal strings \cite{Takahashi,I}. It is useful to introduce the
following $c0$ rapidity functional $\Lambda_{c0}(q) \equiv \sin k
(q)$ where $k(q)$ is the rapidity-momentum functional. The
ground-state values of these rapidity functionals are functions of
the bare-momentum $q$. Such functions are the inverse of the
functions defined in Ref. \cite{V-1}.

Besides an operator representation for the pseudoparticles,
another issue of importance for the introduction of the
pseudofermion description is the definition of suitable Hilbert
subspaces. For instance, an electronic ensemble space is a Hilbert
subspace spanned by all energy eigenstates with the same values
for the electronic numbers $N_{\uparrow}$ and $N_{\downarrow}$. An
important concept is that of {\it CPHS ensemble space}. This is a
subspace spanned by all energy eigenstates with the same values
for the numbers $\{M_{\alpha,\,\pm 1/2}\}$ of $\pm 1/2$ holons
($\alpha =c$) and $\pm 1/2$ spinons ($\alpha =c$) \cite{I,II}. In
general, an electronic ensemble space contains several CPHS
ensemble spaces. Moreover, usually a CPHS ensemble space includes
different {\it CPHS ensemble subspaces}. A CPHS ensemble subspace
is spanned by all energy eigenstates with the same values for the
sets of numbers $N_{c0}$, $\{N_{\alpha\nu}\}$, and
$\{L_{\alpha,\,-1/2}\}$ such that $\alpha =c,s$ and $\nu
=1,2,...$. Here $L_{\alpha,\,\pm 1/2}$ denotes the number of $\pm
1/2$ Yang holons [$\alpha =c$] or $\pm 1/2$ HL spinons [$\alpha
=s$]. (According to the notation of Ref. \cite{II}, CPHS stands
for $c$ pseudoparticle, holon, and spinon.)

Another tool needed for the introduction of the pseudofermion
description is the ground-state normal-ordered pseudoparticle
operational representation. As further discussed below, the
finite-number-electron excitations are contained in a Hilbert subspace
spanned by the initial ground state and excited energy eigenstates
generated from it by processes involving changes in the occupancy
configurations of a finite number of $\alpha\nu$ pseudoparticles,
$-1/2$ Yang holons, and $-1/2$ HL spinons, plus a small finite
density of low-energy and small-momentum $c0$ pseudofermion
(and $s1$ pseudofermion) particle-hole processes for electronic
densities $0<na<1$ (and spin densities $0<ma<na$). Throughout this paper
the symbol $:\hat{O}:$ refers to the ground-state normal-ordered
expression of a general operator $\hat{O}$. Such a ground-state
normal-ordered expression is given by that operator minus its
ground-state expectation value. Let us then introduce the ground-state
normal-ordered $\alpha\nu$ pseudoparticle bare-momentum
distribution operator,

\begin{equation}
:\hat{N}_{\alpha\nu}(q): = \hat{N}_{\alpha\nu}(q) -
N^{0}_{\alpha\nu}(q) \, , \label{noNanop}
\end{equation}
and the $-1/2$ Yang holon ($\alpha =c$) and $-1/2$ HL spinon
($\alpha =s$) ground-state normal-ordered number operator,
$:\hat{L}_{\alpha,\,-1/2}:=\hat{L}_{\alpha,\,-1/2}-L^0_{\alpha,\,-1/2}
=\hat{L}_{\alpha,\,-1/2}$. Here the operators
$\hat{N}_{\alpha\nu}(q)$ and $\hat{L}_{\alpha,\,-1/2}$ are defined
in Eq. (\ref{Nanop}) and in Eq. (39) of Ref. \cite{I},
respectively, $N^{0}_{\alpha\nu} (q)$ is the ground-state
$\alpha\nu$ pseudoparticle distribution function whose expression
is given in Eqs. (C1)-(C3) of the same reference, and $L^0_{\alpha
,\,-1/2}=0$ is the $-1/2$ Yang holon ($\alpha =c$) and $-1/2$ HL
spinon ($\alpha =s$) ground-state number. Note that the
ground-state $c0$ and $s1$ {\it Fermi points} appearing in the
above-mentioned $N^{0}_{c0} (q)$ and $N^{0}_{s1} (q)$ expressions
of Ref. \cite{I} are given by,

\begin{equation}
q^0_{Fc0} = 2k_F \, ; \hspace{0.5cm} q^0_{Fs1} = k_{F\downarrow}
\, . \label{q0Fcs}
\end{equation}
Often one can disregard the $1/L$ corrections and use the
bare-momentum {\it Fermi values} provided in Eq. (\ref{q0Fcs}) and
the limiting bare-momentum values given in Eq. (\ref{qcanGS}).

The eigenvalue equations $:\hat{N}_{\alpha\nu}(q):\vert\psi\rangle
= \Delta N_{\alpha\nu}(q)\vert\psi\rangle$ and
$:\hat{L}_{\alpha,\,-1/2}:\vert\psi\rangle = \Delta
L_{\alpha,\,-1/2}\vert\psi\rangle$ apply to any energy eigenstate
$\vert\psi\rangle$. Here $\Delta N_{\alpha\nu} (q)$ is the
$\alpha\nu$ pseudoparticle bare-momentum distribution function
deviation and $\Delta L_{\alpha ,\,-1/2}$ is the deviation in the
number of $-1/2$ Yang holons ($\alpha =c$) or of $-1/2$ HL spinons
($\alpha =s$). These deviations are given by,

\begin{equation}
\Delta N_{\alpha\nu} (q) \equiv N_{\alpha\nu} (q) -
N^{0}_{\alpha\nu} (q) \, ; \hspace{1cm} \Delta L_{\alpha ,\,-1/2}
\equiv L_{\alpha ,\,-1/2} - L^{0}_{\alpha ,\,-1/2} \, .
\label{DNq}
\end{equation}
Such values refer to excited-energy-eigenstate deviations
relative to the ground-state occupancy configurations described by the
bare-momentum distribution functions and numbers given in Eqs.
(C1)-(C3) of Ref. \cite{I}. For these excited energy eigenstates, the
$\alpha\nu$ pseudoparticle bare-momentum distribution function and
the $-1/2$ Yang holon and $-1/2$ HL spinon numbers read,

\begin{equation}
N_{\alpha\nu} (q) = N^{0}_{\alpha\nu} (q) + \Delta N_{\alpha\nu}
(q) \, ; \hspace{1cm} L_{\alpha ,\,-1/2} = L^{0}_{\alpha ,\,-1/2}
+ \Delta L_{\alpha ,\,-1/2} \, . \label{N0DNq}
\end{equation}
From use of the ground-state distribution and number values given
in Eqs. (C1)-(C3) of Ref. \cite{I}, the operational
relations $:\hat{N}_{c\nu}(q): = \hat{N}_{c\nu}(q)$ for $\nu>0$,
$:\hat{N}_{s\nu}(q): = \hat{N}_{s\nu}(q)$ for $\nu
>1$, and $:\hat{L}_{\alpha,\,-1/2}:=\hat{L}_{\alpha,\,-1/2}$
are founded. Such relations are justified by the absence of the
corresponding quantum objects in the initial ground state.

The transition from the ground state to an excited energy eigenstate
leads to a shift $Q_{\alpha\nu}^0/L=0,\,\pm\pi/L$ in the discrete
bare-momentum value $q_j = [2\pi/L]I^{\alpha\nu}_j$ of Ref. \cite{I}.
That shift is such that,

\begin{eqnarray}
Q_{c0}^0 & = & 0 \, ; \hspace{0.5cm} \sum_{\alpha
=c,\,s}\,\sum_{\nu=1}^{\infty} \Delta N_{\alpha\nu}
\hspace{0.25cm} {\rm even} \, ;  \hspace{1.0cm} Q_{c0}^0=\pm\pi
\, ; \hspace{0.5cm} \sum_{\alpha =c,\,s}\,\sum_{\nu=1}^{\infty}
\Delta
N_{\alpha\nu} \hspace{0.25cm} {\rm odd} \, ; \nonumber \\
Q_{\alpha\nu}^0 & = & 0 \, ; \hspace{0.5cm} \Delta N_{c0}+\Delta
N_{\alpha\nu} \hspace{0.25cm} {\rm even} \, ; \hspace{1.0cm}
Q_{\alpha\nu}^0=\pm\pi \, ; \hspace{0.5cm} \Delta N_{c0}+\Delta
N_{\alpha\nu} \hspace{0.25cm} {\rm odd} \, ; \hspace{0.5cm} \alpha
= c,\,s \, , \hspace{0.25cm} \nu > 0 \, . \label{pic0s1}
\end{eqnarray}
The occurrence of such a bare-momentum shift has effects on the
form of the pseudoparticle operator anticommutators. For instance,
let us consider a $\alpha\nu$ pseudoparticle of bare momentum $q$
and a $\alpha'\nu'$ pseudoparticle of bare momentum $q'$ such that
the bare-momentum values $q$ and $q'$ correspond to an excited
energy eigenstate and the initial ground state, respectively. The
anticommutators involving the creation and/or annihilation
operators of these two pseudoparticles read,

\begin{eqnarray}
\{b^{\dag }_{q,\,\alpha\nu},\,b_{q',\,\alpha'\nu'}\} & = &
\delta_{\alpha,\,\alpha'}\,\delta_{\nu ,\,\nu'}\,\delta_{q,\,q'}
\; \hspace{0.5cm} Q^0_{\alpha\nu}/2 = 0 \nonumber \\
& = & {i\over L}{\delta_{\alpha,\,\alpha'}\,\delta_{\nu
,\,\nu'}\over e^{+i(q-q')/2}\,\sin ([q-q']/2)} \; \hspace{0.5cm}
Q^0_{\alpha\nu}/2 = \pm\pi/2 \, , \label{psGS-S}
\end{eqnarray}
and the anticommutators between two $\alpha\nu$ pseudoparticle
creation or annihilation operators vanish.

\section{THE PSEUDOFERMION DESCRIPTION}

In this section the pseudofermion operational description and the
corresponding pseudoparticle - pseudofermion unitary transformation
are introduced.

\subsection{THE FUNCTIONAL CHARACTER OF THE CANONICAL MOMENTUM}

The $\alpha\nu$ pseudofermion has {\it canonical momentum} ${\bar{q}}_j$ given
by,

\begin{equation}
{\bar{q}}_j = {\bar{q}} (q_j) = q_j + {Q^{\Phi}_{\alpha\nu}(q_j)\over L} =
{2\pi\over L}I^{\alpha\nu}_j + {Q^{\Phi}_{\alpha\nu}  (q_j)\over L} \, ;
\hspace{0.5cm} j=1,2,...,N_{\alpha\nu}^* \, . \label{barqan}
\end{equation}
Here $Q^{\Phi}_{\alpha\nu}(q_j)/L$ is the {\it canonical-momentum shift
functional},

\begin{equation}
{Q^{\Phi}_{\alpha\nu} (q_j)\over L} = {2\pi\over L}
\sum_{\alpha'\nu'}\,
\sum_{j'=1}^{N^*_{\alpha'\nu'}}\,\Phi_{\alpha\nu,\,\alpha'\nu'}(q_j,q_{j'})\,
\Delta N_{\alpha'\nu'}(q_{j'}) \, , \label{qcan1j}
\end{equation}
where the sum $\sum_{\alpha'\nu'}$ runs over all $\alpha'\nu'$
branches with finite pseudofermion occupany in the excited energy
eigenstate. Often in this paper we use the notation $\alpha\nu\neq c0,\,s1$
branches, which refers to all $\alpha\nu$ branches except the $c0$
and $s1$ branches. Moreover, the summations
$\sum_{\alpha\nu}$, $\sum_{\alpha\nu =c0,\,s1}$, and
$\sum_{\alpha\nu\neq c0,\,s1}$
run over all $\alpha\nu$ branches with finite $\alpha\nu$
pseudofermion occupancy in the corresponding state or subspace,
the $c0$ and $s1$ branches only, and all $\alpha\nu$ branches with
finite $\alpha\nu$ pseudofermion occupancy in the corresponding
state or subspace except the $c0$ and $s1$ branches, respectively.

It is shown elsewhere that the function $\pi\Phi_{\alpha\nu,\,\alpha'\nu'}(q,q')$ on the
right-hand side of Eq. (\ref{qcan1j}) is a elementary {\it two-pseudofermion
phase shift} \cite{S-P}. In units of $\pi$ it is given by,

\begin{equation}
\Phi_{\alpha\nu,\,\alpha'\nu'}(q,q') = \bar{\Phi
}_{\alpha\nu,\,\alpha'\nu'}
\left({4t\,\Lambda^{0}_{\alpha\nu}(q)\over U},
{4t\,\Lambda^{0}_{\alpha'\nu'}(q')\over U}\right) \, ,
\label{Phi-barPhi}
\end{equation}
where $\bar{\Phi }_{\alpha\nu,\,\alpha'\nu'} (r ,\,r')$ is the
corresponding rapidity two-pseudofermion phase shift expressed in
terms of the variable $r$ and the ground-state rapidity function
$\Lambda^{0}_{\alpha\nu}(q')$ is defined in terms of its inverse
function in Ref. \cite{V-1}. The rapidity two-pseudofermion
phase shifts are the unique solutions of the integral equations
(\ref{Phis1c})-(\ref{Phisncn}) of Appendix A. The general
expression (\ref{Phi-barPhi}) is valid for densities
$0<na<1$ and $0<ma<na$. The specific two-pseudofermion
phase-shift expressions involving the $c\nu\neq c0$ branches
for $na=1$ (and the $s\nu\neq s1$ branches for $ma=0$)
are studied in Ref. \cite{S-P}.

The functional $Q_{\alpha\nu} (q_j)/L$ such that,

\begin{equation}
Q_{\alpha\nu}(q_j)/2 = Q_{\alpha\nu}^0/2 + Q^{\Phi}_{\alpha\nu} (q_j)/2 \nonumber \\
\, , \label{Qcan1j}
\end{equation}
gives the shift in the discrete canonical-momentum value ${\bar{q}}_j $ that arises
due to the transition from the ground state to an excited energy eigenstate. In turn,
$Q_{\alpha\nu}^0 (q_j)/L$, Eq. (\ref{pic0s1}), gives the corresponding shift in the
discrete bare-momentum value $q_j $ that arises as a result of the same transition.

The momentum shift $Q^{\Phi}_{\alpha\nu} (q_j)/L$ of Eq. (\ref{qcan1j})
is the part of $Q_{\alpha\nu} (q_j)/L$ that refers only to the canonical momentum.
It fully controls the following transformation which relates the $\alpha\nu$
pseudofermion to the corresponding $\alpha\nu$ pseudoparticle,

\begin{equation}
q_j \rightarrow \hspace{0.15cm} {\rm REPLACED} \hspace{0.3cm} {\rm
BY} \rightarrow \hspace{0.3cm} {\bar{q}}_j \, , \label{pfermions}
\end{equation}
where ${\bar{q}}_j$ is the discrete canonical-momentum defined in Eq.
(\ref{barqan}). Note that Eq. (\ref{barqan}) and the transformation (\ref{pfermions})
apply both to the initial ground state and excited energy eigenstates, yet for
the former state $q_j={\bar{q}}_j$ because according to Eq. (\ref{qcan1j}),
$Q^{\Phi}_{\alpha\nu}(q_j)=0$. Indeed, the pseudofermion description refers to the
ground-state normal-ordered 1D Hubbard model. Thus, there is a
specific $\alpha\nu$ pseudofermion description for each initial
ground state. For the latter state the discrete bare momentum $q_j
= [2\pi / L]I^{\alpha\nu}_j$ of Eq. (B1) of Ref. \cite{I} equals
the discrete canonical momentum ${\bar{q}}_j = q_j + Q^{\Phi}_{\alpha\nu}(q_j)/L$.
This justifies the designation {\it bare
momentum} for $q_j$. Thus, for the ground state the $\alpha\nu$
pseudoparticles are identical to the $\alpha\nu$ pseudofermions.
It follows that the ground state is invariant under the
pseudoparticle - pseudofermion transformation and plays
the role of the vacuum of the pseudofermion theory.

The number of $\alpha\nu$ pseudoparticles,
$N_{\alpha\nu}$, equals that of $\alpha\nu$ pseudofermions. Moreover, we introduce the
$\alpha\nu$ canonical-momentum distribution function
${\cal{N}}_{\alpha\nu} ({\bar{q}}_j)$ such that

\begin{equation}
{\cal{N}}_{\alpha\nu} ({\bar{q}}_j)\equiv N_{\alpha\nu}
(q_j({\bar{q}}_j)) \, , \label{NqPF}
\end{equation}
where $N_{\alpha\nu} (q_j)$ stands for the $\alpha\nu$
pseudoparticle bare-momentum distribution function and
$q_j=q_j({\bar{q}}_j)$ is the inverse function of (\ref{barqan}).
The function $N_{\alpha\nu} (q_j)$ is the eigenvalue of the
corresponding bare-momentum number operator (\ref{Nanop}).

A scattering theory for the pseudofermions is introduced in
Ref. \cite{S-P}. In that reference it is found that the invariance under the
electron - rotated-electron unitary transformation of the
$\alpha\nu$ pseudofermions corresponding to $\alpha\nu$ pseudoparticles
created at limiting bare momentum
$q=\pm q_{\alpha\nu}^0$ and belonging to $\alpha\nu\neq c0,\,s1$
branches implies that each of such $c\nu$ pseudofermions (and
$s\nu$ pseudofermions) separates into $2\nu$ independent holons (and
$2\nu$ independent spinons) and a $c\nu$ (and $s\nu$) FP
scattering center. By independent holons and spinons it is meant
those which remain invariant under the electron - rotated-electron
transformation. (The Yang holons and HL spinons are also
independent holons and spinons, respectively.) The above
designation FP stands for {\it Fermi points}. Indeed, it is found
in the same reference that the $c0$ and $s1$ pseudofermion and
pseudofermion hole scatterers feel the created $c\nu$ (and $s\nu$)
FP scattering centers as being $c0$ (and $c0$ and $s1$)
pseudofermion scattering centers at the {\it Fermi points}.

As for the pseudoparticle representation, the
pseudofermion description corresponds to large values of the
Hubbard chain length $L$ and is thus compatible with Takahashi's
string hypothesis \cite{Takahashi,I}. However, while the
pseudoparticle representation corresponds to the whole Hilbert
space, the pseudofermion description and the associated
transformation (\ref{pfermions}) refer to a Hilbert subspace
called {\it pseudofermion subspace} (PS). The PS is spanned by
the initial ground state and the excited energy
eigenstates generated from it by the following types
of processes:\vspace{0.25cm}

(A) - A number $N_{processes}$ of finite-energy and
finite-momentum pseudofermion processes such that $N_{processes}/N_a\rightarrow 0$
as $N_a\rightarrow\infty$, involving creation or
annihilation of $c0$ and $s1$ pseudofermions for all values of
canonical momentum away from the {\it Fermi points} and creation of
pseudofermions belonging to $\alpha\nu\neq c0,\, s1$
branches whose corresponding pseudoparticles have bare-momentum values
obeying the inequality $\vert
q\vert<q_{\alpha\nu}^0$. This can include a number $N_{c0}^{phNF}$ and
$N_{s1}^{phNF}$ of finite-energy and finite-momentum $c0$ and $s1$
pseudofermion particle-hole processes, respectively, such that
$N_{c0}^{phNF}/N_a\rightarrow 0$ and $N_{s1}^{phNF}/N_a\rightarrow
0$ as $N_a\rightarrow\infty$.\vspace{0.25cm}

(B) A number $N_{processes}$ of processes such that
$N_{processes}/N_a\rightarrow 0$ as $N_a\rightarrow\infty$, including
creation of independent $-1/2$ holons and independent $-1/2$ spinons
and zero-energy and finite-momentum processes
which change the number of $c0$ and $s1$ pseudofermions at the
corresponding {\it Fermi points}. (The latter processes transform the
ground-state densely packed $c0$ and $s1$ pseudofermion
canonical-momentum occupancy configuration
into an excited-state densely packed canonical-momentum occupancy
configuration.)\vspace{0.25cm}

(C) - For densities $0<na<1$ and $0<na<ma$, a number
$N_{c0,\,\iota}^{phF}$ and $N_{s1,\,\iota}^{phF}$ of low-energy
and small-momentum elementary $c0$ and $s1$ pseudofermion
particle-hole processes in the vicinity and around the canonical-momentum
values corresponding to the bare-momentum {\it Fermi
points} $\iota\, q^0_{Fc0}$ and $\iota \,q^0_{Fs1}$ of Eq. (\ref{q0Fcs})
where $\iota =\pm 1$, respectively, relative to the excited-state
densely packed $\alpha\nu= c0,\,s1$ pseudofermion canonical-momentum
occupancy configurations generated by the above $c0$ and $s1$ zero-energy
elementary processes (B) and such that $N_{c0,\,\iota}^{phF}/N_a$ and
$N_{s1,\,\iota}^{phF}/N_a$ vanish or remain finite but small as
$N_a\rightarrow\infty$.\vspace{0.25cm}

All PS excited energy eigenstates are generated from the initial
ground state by the processes of types (A)-(C). Thus, the general
PS excited energy eigenstates can be written as $\vert ex\rangle =
G^{\dag} (C)\,G^{\dag} (B)\, G^{\dag} (A) \vert GS\rangle$, where $\vert
GS\rangle$ denotes the initial ground state and $G^{\dag} (A)$, $G^{\dag} (B)$, and
$G^{\dag} (C)$ generate the processes (A), (B), and (C),
respectively. Such generators have simple expressions in terms of
pseudofermion elementary operators \cite{V,V-1}.

For simplicity, densities in the ranges $0<na<1$ and $0<ma<na$ are considered.
The present analysis can be extended to other values of the densities, yet for
the half-filling $na=1$ or zero-magnetization $ma=0$ phases the excitation subspace
is more reduced. The transformation associated with Eq. (\ref{pfermions})
is defined for the PS where it is unitary, as discussed below and in Appendix B.
A crucial point of the pseudofermion theory is that the finite-number-electron
excitations are contained in the PS. By {\it finite-number-electron} operators,
it is meant here operators which can be written as a product of ${\cal N}$ electron
creation and/or annihilation operators such that ${\cal N}/N_a\rightarrow 0$ as
$N_a\rightarrow\infty$. The self-consistency of the pseudofermion
theory is confirmed by the fact that the absolute value of the
spectral-weight matrix elements
between the initial ground state and the PS excited energy eigenstates
obtained by use of the theory is a deceasing function of
the number of finite-energy pseudofermion processes generated by the above
operators $G^{\dag} (A)$ and $G^{\dag} (B)$, vanishing as
$N_{processes}\rightarrow\infty$ \cite{V,V-1}. Such matrix elements are derived by the
pseudofermion dynamical theory reported in Refs. \cite{V,V-1} and are
fully controlled by the functional $Q_{\alpha\nu} (q)/2$ of Eq. (\ref{Qcan1j}).
In turn, the processes (C) can involve a small finite density of
elementary $c0$ and $s1$ pseudofermion particle-hole processes. Indeed, the
residual interactions of the corresponding $c0$ and $s1$ pseudoparticles
vanish for the subspace spanned by the excited states generated by these
processes and thus the $c0$ and $s1$ pseudofermion
energy remains additive in the energy contribution of each elementary
process (C) even if $N_{c0,\,\iota}^{phF}/N_a$ and $N_{s1,\,\iota}^{phF}/N_a$
remain finite but small as $N_a\rightarrow\infty$.

According to Eq. (\ref{Qcan1j}), the processes which generate
the PS excited energy eigenstates from the initial
ground state lead to a collective canonical-momentum shift $Q_{\alpha\nu}
(q_j)/L=Q_{\alpha\nu}^0/L + Q^{\Phi}_{\alpha\nu} (q_j)/L$ for all
the $c0$ pseudofermions and $s1$ pseudofermions of the
initial-state {\it Fermi sea}.
In contrast to the pseudoparticles, the corresponding
pseudofermions have no residual energy interactions. This follows from
the cancelation of the $\alpha\nu$ pseudoparticle residual energy
interactions by the canonical-momentum shift functional $Q^{\Phi}_{\alpha\nu}(q)/L$
of Eq. (\ref{qcan1j}). Such a cancelation is related to
the form of the rapidity functionals $\Lambda_{\alpha\nu}(q)$ and
rapidity-momentum functional $k(q)$ in the PS. Introduction of the
pseudoparticle bare-momentum distribution functions of general
form given in Eq. (\ref{N0DNq}) in the rapidity functional
integral equations (13)-(16) of Ref. \cite{I} and their expansion
in the small deviations (\ref{DNq}), permits explicit solution of
these equations. This procedure leads to expressions for the
rapidity functionals $\Lambda_{\alpha\nu}(q)$ and
rapidity-momentum functional $k(q)$ in terms of the bare-momentum
distribution function deviations introduced in Eq. (\ref{DNq}).
Solution of the above-mentioned integral equations for
distributions of the general form (\ref{N0DNq}) leads to
first-order in the deviations to expressions for the
rapidity-momentum functional and rapidity functionals of the
following form,

\begin{equation}
k (q) = k^0\Bigl({\bar{q}} (q)\Bigr) \, ; \hspace{1cm}
\Lambda_{\alpha\nu}(q) = \Lambda_{\alpha\nu}^0\Bigl({\bar{q}}
(q)\Bigr) \, ; \hspace{1cm} \alpha =c \, ,  \hspace{0.3cm} \nu
=0,1,2,... \, ; \hspace{0.5cm} \alpha =s \, , \hspace{0.3cm} \nu =
1,2,... \, . \label{FL}
\end{equation}
Here ${\bar{q}} (q)$ is the $\alpha\nu$ canonical-momentum functional
given in Eq. (\ref{barqan}) with $q_j$ replaced by the continuum
momentum $q$ and $k^0 (q)$ and $\Lambda_{\alpha\nu}^{0}(q)$ are
the corresponding ground state functions. (These ground-state
functions are fully defined in terms of the corresponding inverse
functions in Ref. \cite{V-1}. Analytical
expressions for $ma=0$ and both $U/t\rightarrow 0$ and $U/t>>1$ are
provided in Ref. \cite{spectral}.)

It is remarkable that in the PS the functionals
$\Lambda_{\alpha\nu}(q)$ and $k(q)$ equal the corresponding
ground-state functions $\Lambda_{\alpha\nu}^0(q)$ and $k^0(q)$,
respectively, with the bare momentum $q$ replaced by the canonical-momentum
functional (\ref{barqan}). This property is behind the lack of
pseudofermion residual energy interactions, as further
discussed in Sec. IV. The canonical-momentum
shift functional (\ref{Qcan1j})
plays a central role in the pseudofermion description of the
finite-number-electron spectral properties.
Indeed, the information recorded in the pseudoparticle
interactions is transferred over to that functional.

The general energy spectrum of the Hamiltonian (\ref{H}) depends
on the quantum object occupancy configurations through the
rapidity and rapidity-momentum functionals and can be written as,

\begin{equation}
E=E_{SO(4)} + \sum_{\alpha =c,\,s}\mu_{\alpha }\,S^{\alpha}_z \, ;
\hspace{0.5cm} E_{SO(4)} = E_H + {U\over 2}\,\Bigl[\,M_c -
2M_{c,\,-1/2} -{N_a\over 2}\Bigr] \, . \label{E}
\end{equation}
where the expression of the energy $E_H$ in terms of the
rapidity-momentum functional $k(q)$ and rapidity functionals
$\Lambda_{\alpha\nu}(q)$ is given in Eq. (20) of Ref. \cite{II}.
As further discussed in Sec. IV, it is this functional character
that is behind the pseudoparticle residual energy interactions.
However, by re-expression of these functionals in terms of the
canonical momentum ${\bar{q}}$, the energy spectrum (\ref{E})
can be written for the PS in terms of pseudofermion canonical-momentum distribution
functions ${\cal{N}}_{c0} ({\bar{q}}_j)$ and ${\cal{N}}_{c\nu}
({\bar{q}}_j)$ as,

\begin{eqnarray}
E & = & -2t \sum_{j=1}^{N_a}\, {\cal{N}}_{c0} ({\bar{q}}_j)\, \cos
k^0({\bar{q}}_j) + 4t \sum_{\nu=1}^{\infty}
\sum_{j=1}^{N^*_{c\nu}}\, {\cal{N}}_{c\nu} ({\bar{q}}_j)\, {\rm
Re}\,\Bigl\{\sqrt{1 - (\Lambda_{c\nu}^0 ({\bar{q}}_j) + i \nu
U/4t)^2}\Bigr\} \nonumber \\ & + & {U\over 2}\Bigl[\,M_c -
\sum_{\nu=1}^{\infty}2\nu\,N_{c\nu} -{N_a\over 2}\Bigr] +
\sum_{\alpha}\mu_{\alpha }S^{\alpha}_z \, . \label{ESS}
\end{eqnarray}
The term $\sum_{\alpha}\mu_{\alpha }S^{\alpha}_z$ is the same as on the
right-hand side of Eq. (\ref{E}) and $N_{c\nu}$
is the number of $c\nu$ pseudofermions.

A crucial point of the $\alpha\nu$ pseudofermion theory is the
replacement of Eq. (\ref{pfermions}) of the bare-momentum $q$ by
the canonical-momentum ${\bar{q}}=q + Q^{\Phi}_{\alpha\nu}(q)/L$. Such a
procedure shows formal similarities with the usual {\it Peierls
substitution}: The pseudofermion, which has no residual energy interactions,
is generated from the corresponding pseudoparticle by the substitution
of the bare-momentum $q$ by the canonical-momentum
${\bar{q}}=q + Q^{\Phi}_{\alpha\nu}(q)/L$. For the PS this
substitution renders the general energy spectrum defined by Eq.
(\ref{E}) and Eq. (20) of Ref. \cite{II} of non-interacting form
for the pseudofermions, as given in Eq. (\ref{ESS}). Indeed, since the
bare-momentum distribution function dependent rapidity functionals
appearing in Eq. (20) of Ref. \cite{II} are replaced by the
corresponding ground-state values $k^0({\bar{q}})$ and
$\Lambda_{c\nu}^0 ({\bar{q}})$, that are independent of the set of
excited-state pseudofermion canonical-momentum distribution functions
$\{{\cal{N}}_{\alpha\nu} (\bar{q})\}$ \cite{spectral}, the energy
(\ref{ESS}) is linear in these functions.

The form of the general energy spectrum (\ref{ESS}) justifies why
the shake-up effects associated with the
functional (\ref{qcan1j}) occur in the case of the pseudofermions in the
canonical momentum instead of in the energy. The dependence of the general
energy spectrum (\ref{ESS}) on that functional occurs
through the canonical momentum in the argument of the ground-state rapidity
and rapidity-momentum functions. Thus, these functions play the
role of non-interacting spectra, since they have the same form
both for the initial ground state and PS excited energy eigenstates. The
shake-up effects associated with the two-pseudofermion phase
shifts are thus {\it felt} by the pseudofermions as mere changes in the
canonical-momentum occupancies, through the canonical-momentum shifts
generated by the ground-state -
excited-energy-eigenstate transitions.

The pseudoparticle bare-momentum $q_j$ description is naturally
provided by the Bethe-ansatz equations \cite{I} within Takahashi's
string hypothesis \cite{Takahashi}. We recall that the
pseudoparticle discrete bare-momentum values $q_j$ are of form
given in Eq. (B1) of  Ref. \cite{I} and according to Eq. (B2) of
the same reference are such that $q_{j+1}-q_j = 2\pi/L$. The
single discrete bare-momentum values $q_j$ are integer multiples
of $2\pi/L$ or of $\pi/L$ \cite{I} and bare-momentum contributions
of order $[1/L]^j$ such that $j>1$ have no physical significance
for the pseudoparticle description: These bare-momentum contributions
must be considered as equaling zero. Importantly,
the same is required for the pseudofermion canonical-momentum discrete
values ${\bar{q}}_j$ given in Eq. (\ref{barqan}). These discrete
values are also at least of the order of $1/L$ and contributions of
order $[1/L]^j$ such that $j>1$ must be considered as equaling
zero. For instance, it is straightforward to find that the
discrete canonical-momentum level separation,

\begin{equation}
{\bar{q}}_{j+1}-{\bar{q}}_j= {2\pi\over L} + {Q^{\Phi}_{\alpha\nu}
(q_{j+1})\over L}- {Q^{\Phi}_{\alpha\nu}  (q_j)\over L} \approx {2\pi\over L} \, ,
\label{differ}
\end{equation}
is such that the second term on the right-hand side of Eq.
(\ref{differ}) is of order $[1/L]^2$, where $\Delta
q_{\alpha\nu}(q)$ is the canonical-momentum shift functional given in Eq.
(\ref{qcan1j}). Thus, up to first order in $1/L$ one finds that
${\bar{q}}_{j+1}-{\bar{q}}_j= 2\pi/L$, as for the corresponding
discrete bare-momentum level separation given in Eq. (B2) of Ref.
\cite{I}. However, this does not imply that to first order in
$1/L$ the pseudofermion canonical-momentum equals the bare-momentum.
Indeed, note that the values of the functional
$Q^{\Phi}_{\alpha\nu}(q_j)/L$ on the right-hand side of Eq. (\ref{barqan})
are of order $1/L$ and play a central role in the control of the
finite-number-electron spectral weight distribution by the non-perturbative
many-electron shake-up effects \cite{V}. What happens is that the level
separation ${\bar{q}}_{j+1}-{\bar{q}}_j= 2\pi/L$ is valid locally
in the discrete canonical-momentum space. By that we mean the following: If
in the present thermodynamic limit two canonical-momentum values
${\bar{q}}_j$ and ${\bar{q}}_{j'}$ differ by a small yet finite
canonical-momentum value $\Delta {\bar{q}}={\bar{q}}_j-{\bar{q}}_{j'}$,
then in general $\Delta {\bar{q}}\neq {2\pi\over L}[j-j']$. In
contrast, for the corresponding bare-momentum values it holds that
$\Delta q= {2\pi\over L}[j-j']$. Therefore, for small but
non-vanishing canonical-momentum separation the difference $Q^{\Phi}_{\alpha\nu}
(q_{j})/L- Q^{\Phi}_{\alpha\nu}  (q_{j'})/L$ is not
anymore of order $[1/L]^2$ and thus has physical significance.

That only discrete canonical-momentum values ${\bar{q}}_j$ of zero and first
order in $1/L$ are physical is an important property of the
pseudofermion theory. Discrete canonical-momentum values of order
$[1/L]^{\cal{N}}$ with ${\cal{N}}>1$ would be generated by
non-linear higher-order terms of the scattering phase-shift functional
$Q^{\Phi}_{\alpha\nu}  (q)/2$ in the pseudofermion canonical-momentum
distribution-function deviations $\Delta N_{\alpha'\nu'}(q')$ on
the right-hand side of Eq. (\ref{Qcan1j}). According to the pseudofermion
scattering studies of Ref. \cite{S-P}, such contributions
would be associated with $({\cal{N}}+1)$-pseudofermion phase
shifts. However, it is shown in that reference that the
corresponding $({\cal{N}}+1)$-pseudofermion $S$ matrix
factorizes into two-pseudofermion $S$ matrices.

\subsection{PSEUDOFERMION OPERATOR ALGEBRA}

The elementary creation and annihilation operators of the
$\alpha\nu$ pseudofermions can be expressed in terms of the
corresponding operators of the $\alpha\nu$ pseudoparticles as
follows,

\begin{equation}
f^{\dag }_{{\bar{q}}_j,\,\alpha\nu} =
{\hat{V}}^{\dag}_{\alpha\nu}\,b^{\dag
}_{q_j,\,\alpha\nu}\,{\hat{V}}_{\alpha\nu} \, ; \hspace{1cm}
f_{{\bar{q}}_j,\,\alpha\nu} =
{\hat{V}}^{\dag}_{\alpha\nu}\,b_{q_j,\,\alpha\nu}\,{\hat{V}}_{\alpha\nu}
\, . \label{f}
\end{equation}
Here ${\hat{V}}_{\alpha\nu}$ is a unitary operator that we call the
{\it $\alpha\nu$ pseudoparticle - pseudofermion
unitary operator}. In Appendix B it is shown that for the PS the operator
${\hat{V}}_{\alpha\nu}$ which obeys Eq. (\ref{f}) is indeed unitary and given
by,

\begin{equation}
{\hat{V}}_{\alpha\nu} =
\exp\bigl\{\sum_{q_j}\,b_{q_j,\,\alpha\nu}^{\dag}[\,b_{q_j+\delta (q_j),\,\alpha\nu}
- b_{q_j,\,\alpha\nu}]\Bigr\} \, ; \hspace{1cm}
\delta (q_j) =Q^{\Phi}_{\alpha\nu}(q_j)/L \, .
\label{Van}
\end{equation}

The canonical-momentum distribution function ${\cal{N}}_{\alpha\nu}
({\bar{q}}_j)$ given in Eq. (\ref{NqPF}) is the eigenvalue of the
operator,

\begin{equation}
{\hat{\cal{N}}}_{\alpha\nu} ({\bar{q}}_j) = f^{\dag
}_{{\bar{q}}_j,\,\alpha\nu}\,f_{{\bar{q}}_j,\,\alpha\nu} \, .
\label{Nqbarop}
\end{equation}
Keeping only the physical momentum contributions that correspond
to terms up to first order in $1/L$, the function
$q_j({\bar{q}}_j)$ appearing in Eq. (\ref{NqPF}) is given by,

\begin{equation}
q_j= q_j ({\bar{q}}_j) = {\bar{q}}_j - \Delta q_{\alpha\nu}
({\bar{q}}_j) = {\bar{q}}_j - {2\pi\over L}
\sum_{\alpha'\nu'}\,
\sum_{j'=1}^{N^*_{\alpha'\nu'}}\,\Phi_{\alpha\nu,\,\alpha'\nu'}^f
({\bar{q}}_j,{\bar{q}}_{j'})\, \Delta {\cal{N}}_{\alpha'\nu'}
({\bar{q}}_{j'}) \, . \label{INVqj}
\end{equation}
(We remind that since the functional (\ref{Qcan1j}) vanishes for the ground state,
$q_j = {\bar{q}}_j$ for that state.) On the
right-hand side of Eq. (\ref{INVqj}),
$\Phi_{\alpha\nu,\,\alpha'\nu'}^f ({\bar{q}},{\bar{q}}')$ is the
canonical-momentum two-pseudfermion phase shift. It is defined as,

\begin{equation}
\Phi_{\alpha\nu,\,\alpha'\nu'}^f ({\bar{q}},{\bar{q}}') =
\Phi_{\alpha\nu,\,\alpha'\nu'}
\Bigl(q({\bar{q}}),q({\bar{q}}')\Bigr) = \bar{\Phi
}_{\alpha\nu,\,\alpha'\nu'}
\left({4t\,\Lambda^{0}_{\alpha\nu}(q({\bar{q}}))\over U},
{4t\,\Lambda^{0}_{\alpha'\nu'}(q({\bar{q}}')))\over U}\right) \, ,
\label{Phi-barPhipf}
\end{equation}
where $q({\bar{q}})$ is the continuum version of the function
(\ref{INVqj}), $\Phi_{\alpha\nu,\,\alpha'\nu'}(q,q')$ is given in
Eq. (\ref{Phi-barPhi}), $\bar{\Phi }_{\alpha\nu,\,\alpha'\nu'} (r
,\,r')$ is the two-pseudofermion phase shift expressed in the
variable $r$ defined by the integral equations
(\ref{Phis1c})-(\ref{Phisncn}) of Appendix A, and
$\Lambda^{0}_{\alpha\nu}(q')$ is defined in terms of its inverse
function in Ref. \cite{V-1}.

Often one replaces the pseudoparticle bare-momentum
summations by integrals and the corresponding discrete
bare-momentum values $q_j$ by a continuum bare-momentum variable
$q$. Since according to Eq. (B2) of Ref. \cite{I}, the difference
$q_{j+1}-q_j = 2\pi/L$ is constant for all values of $j$, the use
of that continuum representation involves the replacement of
$\sum_{j=1}^{N^*_{\alpha\nu}}\equiv\sum_{q=-q_{\alpha\nu
}}^{+q_{\alpha\nu}}$ by ${L\over 2\pi}\int_{-q_{\alpha\nu
}}^{+q_{\alpha\nu}}\,dq$. In the PS, the rapidity functional
$\Lambda_{\alpha\nu}(q)$ and rapidity-momentum functional $k(q)$
equal the corresponding ground-state rapidity function
$\Lambda_{\alpha\nu}^0(q)$ and rapidity-momentum function
$k^0(q)$, respectively, with the bare-momentum $q$ replaced by the
canonical-momentum ${\bar{q}}$. It follows that in the PS the limiting
values of the continuum canonical-momentum ${\bar{q}}$ are given by the
ground-state limiting values $\pm q_{\alpha\nu }^0$ given in Eq.
(\ref{qcanGS}). Thus, to replace the discrete canonical-momentum
values by a continuum canonical-momentum variable ${\bar{q}}$, one must
replace the summations
$\sum_{j=1}^{N^*_{\alpha\nu}}\equiv\sum_{\bar{q}=-q_{\alpha\nu
}^0}^{+q_{\alpha\nu}^0}$ by the integrals ${L\over
2\pi}\int_{-q_{\alpha\nu }^0}^{+q_{\alpha\nu}^0}\,d{\bar{q}}
\,{dq({\bar{q}})\over d{\bar{q}}}$. We then introduce the canonical-momentum
distribution function,

\begin{equation}
{\bar{\cal{N}}}_{\alpha\nu} ({\bar{q}}) = {dq({\bar{q}})\over
d{\bar{q}}}\,{\cal{N}}_{\alpha\nu} ({\bar{q}}) \, . \label{barNq}
\end{equation}
Here,

\begin{equation}
{dq({\bar{q}})\over d{\bar{q}}} = 1 -
\sum_{\alpha'\nu'}\,
\int_{-q_{\alpha'\nu' }^0}^{+q_{\alpha'\nu'}^0}\,d{\bar{q}'}\,
{d\,\Phi_{\alpha\nu,\,\alpha'\nu'}^f({\bar{q}},\,{\bar{q}'})\over
d{\bar{q}}}\,\Delta {\bar{\cal{N}}}_{\alpha'\nu'} ({\bar{q}'}) \,
, \label{INVderq}
\end{equation}
where the function $q = q({\bar{q}})$ is given in Eq.
(\ref{INVqj}) with ${\bar{q}}_j$ replaced by ${\bar{q}}$. The
second term on the right-hand side of this equation is of first
order in the canonical-momentum distribution function deviations. For the
pseudofermion description only canonical-momentum and energy
contributions up to first order in these deviations are physical,
as further discussed below. As a result, for canonical-momentum
distribution function deviations $\Delta
{\bar{\cal{N}}}_{\alpha\nu} ({\bar{q}})$ one can consider that,

\begin{equation}
\Delta {\bar{\cal{N}}}_{\alpha\nu} ({\bar{q}}) = \Delta
{\cal{N}}_{\alpha\nu} ({\bar{q}}) \, , \label{barDNq}
\end{equation}
where in contrast to the case of Eq. (\ref{barNq}) we used
$dq({\bar{q}})/d{\bar{q}}=1$.

Importantly, the $\alpha\nu$ pseudoparticle number operator,

\begin{equation}
{\hat{N}}_{\alpha\nu} = \sum_{j=1}^{N^*_{\alpha\nu}}\,b^{\dag
}_{q_j,\,\alpha\nu}\,b_{q_j,\,\alpha\nu} =
\sum_{q=-q_{\alpha\nu}^0}^{+q_{\alpha\nu}^0}\,b^{\dag
}_{q,\,\alpha\nu}\,b_{q,\,\alpha\nu} = {L\over
2\pi}\int_{-q_{\alpha\nu}^0}^{+q_{\alpha\nu}^0}\,dq \,b^{\dag
}_{q,\,\alpha\nu}\,b_{q,\,\alpha\nu} \, , \label{Nop}
\end{equation}
is invariant under the pseudoparticle - pseudofermion
transformation. It equals the $\alpha\nu$ pseudofermion number
operator,

\begin{equation}
\sum_{j=1}^{N^*_{\alpha\nu}}\,f^{\dag
}_{{\bar{q}}_j,\,\alpha\nu}\,f_{{\bar{q}}_j,\,\alpha\nu} =
\sum_{{\bar{q}}=-q_{\alpha\nu }^0}^{+q_{\alpha\nu}^0}\,f^{\dag
}_{{\bar{q}},\,\alpha\nu}\,f_{{\bar{q}},\,\alpha\nu} = {L\over
2\pi}\int_{-q_{\alpha\nu }^0}^{+q_{\alpha\nu}^0}\,d{\bar{q}}
\,{dq({\bar{q}})\over d{\bar{q}}}\,f^{\dag
}_{{\bar{q}},\,\alpha\nu}\,f_{{\bar{q}},\,\alpha\nu} \, .
\label{Nbarop}
\end{equation}
As shown in Appendix B, this symmetry implies the unitary character
of the $\alpha\nu$ pseudoparticle - pseudofermion
operator (\ref{Van}).

Moreover, the $c\nu$ pseudoparticle charge and the spin and
$\eta$-spin values
found and provided in Ref. \cite{I} are also invariant under the
above transformation. The same occurs for the $s\nu$
pseudoparticle spin and spin-projection values
given in that reference. The pseudoparticle - pseudofermion
transformation also leaves invariant the $\pm 1/2$ Yang holons and
$\pm 1/2$ HL spinons. The $\pm 1/2$ holon (and $\pm 1/2$ spinon)
composite character of the $c\nu\neq c0$ pseudoparticles (and $s\nu$
pseudoparticles) also remains invariant under that transformation.
It follows that the $c\nu\neq c0$ pseudofermions (and $s\nu$
pseudofermions) are $\eta$-spin zero (and spin zero)
composite objects of an equal number $\nu=1,2,...$ of $-1/2$
holons and $+1/2$ holons (and $-1/2$ spinons and $+1/2$ spinons).
Thus, by combining Eqs. (24) and (30) of Ref. \cite{I} the $\pm
1/2$ holon ($\alpha =c$) and $\pm 1/2$ spinon ($\alpha =s$) number
operators ${\hat{M}}_{\alpha,\,\pm 1/2}$ can be written in terms
of pseudofermion operators as follows,

\begin{equation}
{\hat{M}}_{\alpha,\,\pm 1/2} = {\hat{L}}_{\alpha,\,\pm 1/2} +
\sum_{\nu=1}^{\infty}\,\sum_{q=-q_{\alpha\nu }^0}^{+q_{\alpha\nu
}^0}\nu\,{\hat{\cal{N}}}_{\alpha\nu} (\bar{q}) \, . \label{Mopf}
\end{equation}
Here the pseudfermion canonical-momentum distribution operator
${\hat{\cal{N}}}_{\alpha\nu} (\bar{q})$ is given in Eq.
(\ref{Nqbarop}) and the operator ${\hat{L}}_{\alpha,\,\pm 1/2}$ is
the $\pm 1/2$ Yang holon ($\alpha =c$) and $\pm 1/2$ HL spinon
($\alpha =s$) number operator provided in Eq. (39) of Ref.
\cite{I}. Thus, for the PS all results reported in Ref. \cite{I} concerning
pseudoparticle charge and spin transport are also valid for the
corresponding pseudofermions. For instance, for finite values of
$U/t$ the transport of charge (and spin) is associated with the
$c0$ pseudofermion and $c\nu$ pseudofermion quantum charge fluids
(and $s\nu$ pseudofermion quantum spin fluids).

We recall that the bare-momentum $q$ is the conjugate of the
spatial coordinate $x_j =a_{\alpha\nu}\,j$ associated with the
effective $\alpha\nu$ lattice, where $j=1,2,...,N^*_{\alpha\nu}$.
As for the charge (or spin) carried by the
pseudoparticles and of their composite character in terms of
chargeaons and antichargeons \cite{I}, $\pm 1/2$ holons, or $\pm
1/2$ spinons, also the effective $\alpha\nu$ lattice remains
invariant under the $\alpha\nu$ pseudoparticle - $\alpha\nu$
pseudofermion unitary transformation. Indeed, the functional
$Q^{\Phi}_{\alpha\nu}(q)/L$ which controls the
pseudoparticle - pseudofermion transformation (\ref{pfermions})
does not affect the underlying effective $\alpha\nu$ lattice. As shown
in Ref. \cite{S-P}, for the $\alpha\nu = c0,\, s1$ branches with finite
occupancy in the initial ground state, that momentum-shift functional
just imposes a twisted boundary condition.

As for the case of the pseudoparticles, it is useful to introduce
the local $\alpha\nu$ pseudofermion creation operator $f^{\dag
}_{x_j,\,\alpha\nu}$ and annihilation operator
$f_{x_j,\,\alpha\nu}$. These operators are related to the
operators $f^{\dag }_{\bar{q},\,\alpha\nu}$ and
$f_{\bar{q},\,\alpha\nu}$, respectively, obtained from the
corresponding pseudoparticle operators through the relations given
in Eq. (\ref{f}), as follows,

\begin{equation}
f^{\dag }_{\bar{q},\,\alpha\nu} =
{1\over\sqrt{N^*_{\alpha\nu}}}\sum_{j=1}^{N^*_{\alpha\nu}}e^{i\bar{q}\,x_j}\,
f^{\dag }_{x_j,\,\alpha\nu} \, ; \hspace{1cm}
f_{\bar{q},\,\alpha\nu} =
{1\over\sqrt{N^*_{\alpha\nu}}}\sum_{j=1}^{N^*_{\alpha\nu}}e^{-i\bar{q}\,x_j}\,
f_{x_j,\,\alpha\nu} \, ,  \label{elopf}
\end{equation}
where the summations refer to the sites of the effective
$\alpha\nu$ lattice. The local $\alpha\nu$ pseudofermion creation
(and annihilation) operator $f^{\dag }_{x_j,\,\alpha\nu}$ (and
$f_{x_j,\,\alpha\nu}$) creates (and annihilates) a $\alpha\nu$
pseudofermion at the effective $\alpha\nu$ lattice site of spatial
coordinate $x_j =a_{\alpha\nu}^0\,j$, where
$j=1,2,...,N^*_{\alpha\nu}$ and $a_{\alpha\nu}^0$ is the effective
$\alpha\nu$ lattice constant given in Eq. (\ref{acanGS}). (For the
PS and except for $1/L$ corrections we can consider that the
effective $\alpha\nu$ lattice constants are the ground-state
constants $a_{\alpha\nu}^0$.) Thus, the conjugate variable of the
canonical momentum ${\bar{q}}_j$ of the $\alpha\nu$ pseudofermion branch is
the space coordinate $x_j$ of the corresponding effective
$\alpha\nu$ lattice. The local $\alpha\nu$ pseudoparticles and
corresponding local $\alpha\nu$ pseudofermions have the same
effective $\alpha\nu$ lattice. It follows that the local
pseudoparticle and local pseudofermion site distribution
configurations which describe the ground state and the PS excited
energy eigenstates are the same. (These configurations are expressed in terms
of rotated-electron site distribution configurations in Ref.
\cite{IIIb}.)

While the local $\alpha\nu$ pseudoparticles and corresponding
local $\alpha\nu$ pseudofermions {\it live} in the same effective
$\alpha\nu$ lattice, the values of the set of discrete
bare-momentum values $\{q_j\}$ and canonical-momentum values
$\{{\bar{q}}_j\}$ such that $j=1,2,...,N^*_{\alpha\nu}$ are
different and related by Eq. (\ref{INVqj}). There is an one-to-one
relation between these two sets of discrete values, which keep the
same order because there is no level crossing. This property
follows from the values of the discrete bare-momentum and canonical-momentum
separation given in Eq. (B2) of Ref. \cite{I} and Eq.
(\ref{differ}), respectively.

Finally, we consider the anticommutation relations of the
pseudofermion operators. It is confirmed in Refs. \cite{V,V-1} that
such relations play a major role in the evaluation of finite-number-electron
matrix elements between energy eigenstates. Let us consider the general
situation when the canonical momenta ${\bar{q}}$ and ${\bar{q}}'$ of the
operators $f^{\dag }_{{\bar{q}},\,\alpha\nu}$ and
$f_{{\bar{q}}',\,\alpha\nu}$, respectively, correspond to
different CPHS ensemble subspaces. The anticommutator $\{f^{\dag
}_{{\bar{q}},\,\alpha\nu},\,f_{{\bar{q}}',\,\alpha'\nu'}\}$ can be
expressed in terms of the local-pseudofermion anticommutators
$\{f^{\dag }_{x_j,\,\alpha\nu},\,f_{x_{j'},\,\alpha'\nu'}\}$
associated with spatial coordinates $x_j$ and $x_{j'}$ of the
effective $\alpha\nu$ and $\alpha'\nu'$ lattices, respectively, as
follows,

\begin{equation}
\{f^{\dag
}_{{\bar{q}},\,\alpha\nu},\,f_{{\bar{q}}',\,\alpha'\nu'}\} =
{1\over
{\sqrt{N^*_{\alpha\nu}N^*_{\alpha'\nu'}}}}\,
\sum_{j=1}^{N^*_{\alpha\nu}}\,\sum_{j'=1}^{N^*_{\alpha'\nu'}}\,
e^{i({\bar{q}}\,x_j-{\bar{q}}'\,x_{j'})}\, \{f^{\dag
}_{x_j,\,\alpha\nu},\,f_{x_{j'},\,\alpha'\nu'}\} \, .
\label{pfacrjj}
\end{equation}
The momentum operator, $\hat{P} = \sum_{\sigma=\uparrow
,\,\downarrow }\sum_{k}\, \hat{N}_{\sigma} (k)\, k$, which is the
generator for the spatial translations, commutes with the unitary
electron - rotated-electron transformation. Thus, it has the same
expression, $\hat{P} = \tilde{P}=\sum_{\sigma=\uparrow,\,\downarrow }\sum_{k}\,
\tilde{N}_{\sigma} (k)\, k$, where $\hat{N}_{\sigma} (k) =
{\tilde{c}}_{k,\,\sigma }^{\dagger }\,{\tilde{c}}_{k,\,\sigma }$,
in terms of creation and annihilation rotated-electron operators.
It follows that the electronic lattice remains invariant under
such a transformation and, therefore, the rotated-electron
lattice and corresponding lattice constant equal those of the original
electrons. Furthermore, there is a direct relation between the latter
lattice and the effective $\alpha\nu$ lattice populated by the local
$\alpha\nu$ pseudofermions associated with the operators
$f^{\dag }_{x_j,\,\alpha\nu}$ and $f_{x_{j'},\,\alpha'\nu'}$. The above
invariance implies that the local-pseudofermion anticommutators
$\{f^{\dag }_{x_j,\,\alpha\nu},\,f_{x_{j'},\,\alpha'\nu'}\}$ have
simple expressions. However, under the $j$ and $j'$ summations
of Eq. (\ref{pfacrjj}) the the exotic functional character of the
canonical-momentum values leads to the following unusual algebra
for the $\alpha\nu$ pseudofermion operators,

\begin{equation}
\{f^{\dag }_{{\bar{q}},\,\alpha\nu},\,f_{{\bar{q}}',\,\alpha'\nu'}\}
= \delta_{\alpha\nu,\,\alpha'\nu'}\,{1\over
N^*_{\alpha\nu}}\,e^{-i({\bar{q}}-{\bar{q}}')\,a/
2}\,e^{i(Q_{\alpha\nu}(q)-{Q'}_{\alpha\nu}
(q'))/2}\,{\sin\Bigl([Q_{\alpha\nu} (q)-{Q'}_{\alpha\nu}(q')]/
2\Bigr)\over\sin ([{\bar{q}}-{\bar{q}}']\,a/2)} \, , \label{pfacr}
\end{equation}
and the anticommutators between two $\alpha\nu$ pseudofermion
creation or annihilation operators vanish. Here the
values $Q_{\alpha\nu}(q)/2$ and ${Q'}_{\alpha\nu}(q')/2$
of the canonical-momentum shift functional (\ref{Qcan1j}) refer to the CPHS
ensemble subspaces which the canonical momenta ${\bar{q}}$ and ${\bar{q}}'$
correspond to, respectively.

A case of particular importance is when the CPHS ensemble
subspaces associated with the canonical momentum ${\bar{q}}'$ is that of the
initial ground state. In that case ${Q'}_{\alpha\nu}(q')/2=0$ for
the ground-state CPHS ensemble subspace and thus the
anticommutation relation (\ref{pfacr}) simplifies to,

\begin{equation}
\{f^{\dag }_{{\bar{q}},\,\alpha\nu},\,f_{{\bar{q}}',\,\alpha'\nu'}\}
= \delta_{\alpha\nu,\,\alpha'\nu'}\,{1\over
N^*_{\alpha\nu}}\,e^{-i({\bar{q}}-{\bar{q}}')\,a/
2}\,e^{iQ_{\alpha\nu}(q)/2}\,{\sin\Bigl(Q_{\alpha\nu} (q)/
2\Bigr)\over\sin ([{\bar{q}}-{\bar{q}}']\,a/2)} \, . \label{pfacrGS}
\end{equation}
Note that if $\sin (Q_{\alpha\nu} (q)/2)$ would vanish the
anticommutation relation (\ref{pfacrGS}) would be the usual one,
$\{f^{\dag
}_{{\bar{q}},\,\alpha\nu},\,f_{{\bar{q}}',\,\alpha'\nu'}\} =
\delta_{\alpha,\,\alpha'}\,\delta_{\nu
,\,\nu'}\,\delta_{{\bar{q}},\,{\bar{q}}'}$. In contrast, in our case
that quantity has in general finite values. Indeed, under nearly all
ground-state - excited-energy-eigenstate transitions there is
a canonical-momentum shift which results from non-perturbative
shake-up effects.

For the $c\nu\neq c0$ (and $s\nu\neq s1$) pseudofermion branches the
anticommutation relations (\ref{pfacr}) and (\ref{pfacrGS}) refer to
electronic densities $na<1$ (and spin densities $m>0$) such that
$N^*_{c\nu}/N_a$ (and $N^*_{s\nu}/N_a$) are finite. Fortunately, for
the scattering properties only the $c0$ and $s1$ pseudofermion
anticommutation relations (\ref{pfacrGS}) are needed \cite{V} and
thus one can study these properties for densities such that
$0< na\leq 1$ and $0< ma\leq na$.

Finally, comparison of Eqs. (\ref{psGS-S}) and (\ref{pfacrGS})
for the $\alpha\nu$ pseudoparticle and $\alpha\nu$ pseudofermion
anticommutators, respectively, reveals that the absence of pseudofermion residual
energy interactions implies a more complicated functional character
for the $\alpha\nu$ pseudofermion anticommutator. Indeed,
for the pseudofermion description the whole information contained
in the pseudoparticle residual interactions is transferred over to the
pseudofermion anticommutator (\ref{pfacrGS}), through the unitary
transformation described by Eqs. (\ref{pfermions}) and (\ref{f}).

\section{THE PSEUDOFERMION ENERGY AND MOMENTUM SPECTRA AND THE
WAVE-FUNCTION FACTORIZATION OF THE NORMAL-ORDERED 1D HUBBARD
MODEL}

In this section, we find that the description of the quantum
problem in terms of the pseudofermions leads in
the thermodynamic limit to a wave-function factorization for the
PS excited energy eigenstates where the excitations generated by application
onto the ground state of finite-number-electron operators are contained.
Such a factorization refers to the ground-state normal-ordered 1D
Hubbard model.

\subsection{THE PSEUDOFERMION ENERGY AND MOMENTUM SPECTRA}

In the PS the energy spectrum is of the form given in Eq.
(\ref{ESS}). The processes which generate the PS excited energy
eigenstates from the ground state can be associated with two
virtual excitations: (1) a finite number of elementary processes (A)
and (B) followed by small-momentum and low-energy $c0$ pseudofermion
and $s1$ pseudofermion particle-hole processes (C); (2) a collective
canonical-momentum shift $Q_{\alpha\nu} (q)/L$ for all $\alpha\nu$
pseudofermions and $\alpha\nu$ pseudofermion holes with finite
pseudofermion occupancy in the excited energy eigenstate. The
corresponding energy deviation spectrum is derived from the general
PS energy spectrum given in Eq. (\ref{ESS}). Such a energy deviation
spectrum corresponds to the ground-state normal-ordered 1D Hubbard
model and reads,

\begin{equation}
\Delta E = \omega_0 + \sum_{j=1}^{N_a}\,\Delta {\cal{N}}_{c0}
({\bar{q}}_j) \,\epsilon_{c0} ({\bar{q}}_j) +
\sum_{j=1}^{N^*_{s1}}\,\Delta {\cal{N}}_{s1} ({\bar{q}}_j)
\,\epsilon_{s1}({\bar{q}}_j) + \sum_{\alpha\nu\neq c0,\,s1}\,
\sum_{j=1}^{N^*_{\alpha\nu}}\,\Delta {\cal{N}}_{\alpha\nu}
({\bar{q}}_j)\,\epsilon^0_{\alpha\nu}({\bar{q}}) \, , \label{E1pf}
\end{equation}
where the energy parameter $\omega_0$ is given by,

\begin{equation}
\omega_0 = 2\mu\, \Delta M_{c,\,-1/2} + 2\mu_0\,H\, [\Delta
M_{s,\,-1/2}-\Delta N_{s1}] \, . \label{om0}
\end{equation}
Here, $\Delta M_{\alpha ,\,-1/2}$ are the deviations in the
numbers of $-1/2$ holons ($\alpha =c$) and of $-1/2$ spinons
($\alpha =s$) and $\Delta N_{s1}$ is the
deviation in the number of $s1$ pseudofermions. The energy
deviation spectrum (\ref{E1pf}) is additive in the $-1/2$ holon, $-1/2$
spinons, and $\alpha\nu$ pseudofermion energies. On the right-hand
side of Eq. (\ref{E1pf}), the functions $\epsilon_{c0}
({\bar{q}})$, $\epsilon_{s1}({\bar{q}})$, and
$\epsilon^0_{\alpha\nu}({\bar{q}})$ are the pseudofermion energy
bands defined in Eqs. (C.15)-(C.21) of Ref. \cite{I}. The
zero-energy levels relative to the initial ground-state
energy of these dispersions are such that,

\begin{equation}
\epsilon_{c0} (\pm 2k_F) = \epsilon_{s1} (\pm k_{F\downarrow})=
\epsilon_{c\nu}^0 (\pm [\pi/a -2k_F])=\epsilon_{s\nu}^0 (\pm
[k_{F\uparrow}-k_{F\downarrow}])=0 \, . \label{eplev0}
\end{equation}
Note that the above pseudofermion energy bands equal the
corresponding pseudoparticle energy bands \cite{I,II} provided
that the bare momentum $q$ is replaced by the canonical momentum $\bar{q}$.
The latter bands are plotted in Figs. 6 to 9 of Ref. \cite{II}
for $ma=0$.

In Ref. \cite{S-P} it is found that for the PS both the scattering-less
phase shift $Q^0_{\alpha\nu}/2$ and the scattering phase shift
$Q^{\Phi}_{\alpha\nu} (q_j)/2$ conserve the total energy. Thus,
the virtual-excitation (2) energy spectrum
vanishes and the general deviation-linear energy spectrum
(\ref{E1pf}) amounts to the contributions from the excitation (1).
The latter PS excitation involves changes in the occupancy configurations
of a finite number of quantum objects generated by the elementary processes
(A) and (B) plus the small-momentum and low-energy excitations
generated by the $c0$ and $s1$ pseudofermion particle-hole elementary
processes (C). The excitation (2) involves
a collective canonical-momentum shift of all the ground-state $c0$
pseudofermions and $s1$ pseudofermions. In the thermodynamic limit, the
value of the corresponding ground-state numbers $N_{c0}=N$ and
$N_{s1}=N_{\downarrow}$ approaches infinity. Thus, the
self-consistency of the pseudofermion theory
implies that the energy of the excitation (2) vanishes in that
limit, so that the total energy is additive in the corresponding
pseudofermion energies. This requirement is fulfilled because
both the pseudofermion scattering events and the scattering-less
phase shift $Q^0_{\alpha\nu}/2$ conserve the total energy \cite{S-P}.
In contrast, the elementary processes (C) do not involve all $c0$
and $s1$ pseudofermions, yet the corresponding densities
$N_{c0,\,\iota}^{phF}/N_a$ and $N_{s1,\,\iota}^{phF}/N_a$ can
reach finite but small values as $N_a\rightarrow\infty$. However, we
find below that in spite of these values the corresponding energy
spectrum remains linear in the pseudofermion particle-hole
numbers $N_{c0,\,\iota}^{phF}$ and $N_{s1,\,\iota}^{phF}$.

The general PS energy spectrum defined by Eq. (\ref{ESS}) has no
pseudofermion residual-interaction energy terms for it is
linear in the pseudofermion canonical-momentum distribution functions. In
contrast, when expressed in terms of pseudoparticle bare-momentum
distribution functions it includes residual-interaction terms. In the
latter case it is defined by Eq. (\ref{E}) and Eq. (20) of Ref.
\cite{II}. Such residual-interaction energy terms arise from the
dependence on the pseudoparticle bare-momentum distribution functions
of the rapidity functionals $k (q)$ and $\Lambda_{c\nu} (q)$ on the
right-hand side of the energy expression (20) of Ref. \cite{II}.
That dependence is defined by the integral equations (13)-(16)
of Ref. \cite{I}. It is the surprising property that in the PS the
rapidity functionals $k (q)$ and $\Lambda_{c\nu} (q)$ obey the
relations given in Eq. (\ref{FL}), where $k^0 (q)$ and
$\Lambda_{c\nu}^{0}(q)$ are the corresponding ground-state values,
that is behind the absence of pseudofermion residual-interaction
energy terms in expression (\ref{ESS}), as discussed in Sec. III. Indeed,
reexpression of the general energy spectrum defined by Eq. (\ref{E})
and Eq. (20) of Ref. \cite{II} in terms of the canonical momentum
$\bar{q}$ leads to the replacement of the functionals $k (q)$ and
$\Lambda_{c\nu} (q)$ by their ground-state values $k^0 (q)$ and
$\Lambda_{c\nu}^{0}(q)$, respectively. This {\it Peierls substitution}
procedure leads to the PS general energy spectrum (\ref{ESS}),
which is linear in the canonical-momentum distribution functions.
The non-interacting character of that energy spectrum requires that
in the PS the general energy-deviation spectrum must also be linear
in the pseudofermion canonical-momentum distribution-function deviations
$\Delta {\cal{N}}_{\alpha\nu} ({\bar{q}})$ of Eq. (\ref{barDNq}). Indeed,
if the energy-deviation spectrum involved non-linear contributions in the
pseudofermion canonical-momentum distribution-function deviations, these
quantum objects would have residual interactions. This implies that for the
pseudofermion theory the higher-order energy contributions in the
pseudofermion canonical-momentum distribution-function deviations have no
physical meaning. This is consistent with full information about
the spectral properties being retained in the
functional $Q^{\Phi}_{\alpha\nu}(q)/L$ of Eq.
(\ref{qcan1j}). Therefore, in order to check the consistency of
the pseudofermion theory, one should confirm that in the
thermodynamic limit only the energy-deviation spectrum
(\ref{E1pf}) is finite, and thus that the energy terms originated
by higher-order contributions in the pseudofermion canonical-momentum
distribution-function deviations vanish in that limit.

In the following we confirm that energy contributions of order
equal or larger than two in the canonical-momentum
distribution-function deviations vanish in the thermodynamic
limit. For that we use the relation ${\cal{N}}_{\alpha\nu}
({\bar{q}}_j) = N_{\alpha\nu} (q_j({\bar{q}}_j))$ given in Eq.
(\ref{NqPF}). Such a relation implies that for the PS the energy
contributions originated by processes of the same order in the
canonical-momentum and bare-momentum distribution-function
deviations differ only by the contributions of the collective
shift $Q_{\alpha\nu}^{\Phi} (q)/L$ of the
overall canonical-momentum shift $Q_{\alpha\nu} (q)/L$ associated with the
excitation (2). However, since the energy contributions of the excitation (2)
vanish \cite{S-P}, the above two energy contributions
are identical, referring to the
excitation (1) only. Thus, if the energy contributions in the
bare-momentum distribution-function deviations of order
equal or larger than two vanish in the thermodynamic limit, the
same occurs for the corresponding contributions in the
canonical-momentum distribution-function deviations. Moreover, we
also confirm that when the ratios $N_{c0,\,\iota}^{phF}/N_a$ and
$N_{s1,\,\iota}^{phF}/N_a$ remain finite but small as
$N_a\rightarrow\infty$, the energy spectrum associated with
the elementary processes (C) is linear in the numbers
$N_{c0,\,\iota}^{phF}$ and $N_{s1,\,\iota}^{phF}$.

For the PS both the bare-momentum and the canonical-momentum
distribution-function deviations describe the elementary processes
(A)-(C) which generate the excited energy eigenstates
from the ground state. The processes (A) and (B) of the above
excitation (1) are in finite number. Thus, in the thermodynamic limit these
processes involve a vanishing density of quantum objects. Let us
consider a more extended Hilbert space spanned by excited energy
eigenstates involving a small but vanishing or finite density of occupied
bare-momentum values for the bare-momentum distribution function
deviations generated by the elementary processes of type (A)
and (B). In this case, one can derive an energy functional
in terms of the bare-momentum distribution function deviations
$\Delta N_{c0}(q)$, $\{\Delta N_{c\nu}(q)\}$, and $\{\Delta
N_{s\nu}(q)\}$ of Eq. (\ref{DNq}) where $\nu =1,2,...$. This is
achieved by solution of the integral equations (13)-(16) of Ref.
\cite{I} for bare-momentum distribution functions of the general
form (\ref{N0DNq}). Use of the obtained rapidity functional
expressions in the general energy spectrum defined by Eq.
(\ref{E}) and Eq. (20) of Ref. \cite{II} leads to a general energy
functional of the following form,

\begin{equation}
\Delta E = \sum_{i=1}^{\infty}\Delta E_i \, , \label{EF}
\end{equation}
where the index $i=1,2,...$ refers to the order in the above
bare-momentum distribution function deviations. (For
the pseudoparticle description, the terms of order
$i$ larger than one describe the residual interactions of the
pseudoparticles.) The first and second-order terms on the
right-hand side of Eq. (\ref{EF}) are of the following general
form,

\begin{equation}
\Delta E_1 = \omega_0 + \sum_{q=-q_{c0}^0}^{+q_{c0}^0}
\,\epsilon_{c0} (q)\Delta N_{c0} (q) +
\sum_{q=-q_{s1}^0}^{+q_{s1}^0}\epsilon_{s1}(q)\,\Delta N_{s1}(q) +
\sum_{\alpha\nu\neq c0,\,s1}\,\sum_{q=-\Delta
q_{\alpha\nu}^0}^{+\Delta q_{\alpha\nu}^0}
\epsilon^0_{\alpha\nu}(q)\,\Delta N_{\alpha\nu}(q) \, , \label{E1}
\end{equation}
and

\begin{eqnarray}
\Delta E_2 = {1\over
L}\Bigl\{\,\sum_{\alpha\nu}\,\sum_{q=-q^0_{\alpha\nu}}^{+q^0_{\alpha\nu}}
v_{\alpha\nu}(q)\,Q^{\Phi}_{\alpha\nu}(q)\,\Delta N_{\alpha\nu}(q)
+ {1\over 4\pi}\sum_{\alpha\nu =c0,\,s1} v_{\alpha\nu} \sum_{j=\pm
1}\Bigl[Q^{\Phi}_{\alpha\nu}(jq^0_{F\alpha\nu})\Bigr]^2\Bigr\} \,
, \label{E2}
\end{eqnarray}
respectively. Here $\Delta N_{c0} (q)$, $\Delta N_{c\nu} (q)$,
and $\Delta N_{s\nu} (q)$ are the bare-momentum distribution function
deviations given in Eq. (\ref{DNq}) and $\omega_0$ is the energy functional
(\ref{om0}). This functional is linear in the deviations $\Delta M_{c
,\,-1/2}$ $\Delta M_{s ,\,-1/2}$, and $\Delta N_{s1}$.
(The $s1$ pseudofermion number deviation of Eq.
(\ref{om0}) equals the corresponding $s1$ pseudoparticle number
deviation.) Note that the linearity in the $-1/2$ Yang holon and
$-1/2$ HL spinon number deviations holds for all excited energy
eigenstates independently of whether the values of these deviations
are small or large. That behavior follows from the non-interacting
character of these quantum objects \cite{II}. The coefficients of
the $i=1$ linear terms are the $\alpha\nu$ energy
bands $\epsilon_{c0}(q)$, $\epsilon_{s1}(q)$, and
$\epsilon_{\alpha\nu}^0(q)$. (These bands equal the pseudofermion
energy bands with the canonical-momentum $\bar{q}$ replaced by the
bare-momentum $q$.) Importantly, note that the coefficients of the $i=2$
pseudoparticle residual-interaction energy quadratic terms (\ref{E2}) involve the
canonical-momentum-shift functional $Q^{\Phi}_{\alpha\nu}(q)/L$ defined
in Eq. (\ref{qcan1j}), besides the $\alpha\nu$ group velocities $v_{\alpha\nu} (q)=\partial
\epsilon_{\alpha\nu} (q)/\partial q$ and $v_{\alpha\nu}=
v_{\alpha\nu} (q^0_{F\alpha\nu})$. For the pseudoparticles such
a functional appears in the residual-interaction energy terms, whereas
for the pseudofermions it is transferred over to the canonical momentum
(\ref{barqan}) by the unitary transformation given in Eqs. (\ref{pfermions})
and (\ref{f}). By use of Eq. (\ref{qcan1j}) in
the first expression of Eq. (\ref{E2}), the $i=2$ energy $\Delta
E_2$ can be expressed in terms of suitable pseudoparticle $f$
functions $f_{\alpha\nu,\,\alpha'\nu'}(q,q')$ \cite{Carmelo92,S-P}.
This leads to $\alpha\nu$, $\alpha'\nu'$, $q$, and $q'$ summations
of energy terms of the form $[1/2L]\,f_{\alpha\nu,\,
\alpha'\nu'}(q,q')\,\Delta N_{\alpha\nu}(q)\,\Delta
N_{\alpha',\,\nu'}(q')$. An important property is that the
coefficients of all energy terms $E_i$ of expression (\ref{EF}) of
order $i>1$ involve only the group velocities, the
two-pseudofermion phase shifts, and/or bare-momentum derivatives
of these functions. For instance, note that $E_2$ involves the
velocities and the functional $Q^{\Phi}_{\alpha\nu}(q)/2$ of Eq.
(\ref{qcan1j}). This property is related to the two-pseudofermion
reducibility of the $({\cal{N}}+1)$-pseudofermion $S$ matrix \cite{S-P}.
Therefore, the two-pseudofermion phase shifts control the energy expansion
(\ref{EF}) for all deviation orders.

First, we note that although the numbers $N_{c0,\,\iota}^{phF}$ and
$N_{s1,\,\iota}^{phF}$ of small-momentum and low-energy elementary
$c0$ and $s1$ pseudofermion particle-hole processes (C) are such that
the corresponding ratios $N_{c0,\,\iota}^{phF}/N_a$ and
$N_{s1,\,\iota}^{phF}/N_a$ can reach finite but small values
as $N_a\rightarrow\infty$,
such processes do not contribute to the value of the functional
$Q^{\Phi}_{\alpha\nu} (q_j)/2$. Indeed, the "particle" contributions to
that functional are exactly canceled by the corresponding "hole"
contributions \cite{V-1}. This behavior follows from the vanishing
of the $c0$ and $s1$ pseudoparticle residual interactions in
the subspace spanned by the excited states generated by the
elementary processes (C), as confirmed by the form of the
energy $E_2$, Eq. (\ref{E2}), and of the remaining energy terms $E_i$ of
expression (\ref{EF}) of order $i>2$, which vanish when
$Q^{\Phi}_{\alpha\nu} (q_j)/2=0$. It follows that the $c0$ and $s1$
pseudofermion particle-hole processes (C) only contribute to
the leading-order energy term $E_1$, Eq. (\ref{E1}), of
expression (\ref{EF}). Thus, the energy spectrum of the
small-momentum and low-energy elementary $c0$ and $s1$ pseudofermion
particle-hole excitations generated by the elementary processes
(C) remains linear in the number of these processes even
for a small finite density of such processes.

Second, we address the issues related to the energy spectrum associated
with the elementary processes (A) and (B). The energy (\ref{EF}) is an
expansion in the bare-momentum distribution function deviations. For
excited energy eigenstates whose occupancy configurations are generated
by processes similar to the elementary processes (A) and (B) but
involving a small finite density of occupied bare-momentum values for the
bare-momentum distribution function deviations, all energy terms
of deviation order $i=1,2,...$
are of the same order $[1/L]^{-1}=L$ in $1/L$. However, here we
are interested in the PS where the finite-number-electron excitations are
contained. The PS excited-state deviations associated with the occupancy
configurations generated by the elementary processes (A) and (B)
involve a vanishing density of pseudofermions, $-1/2$ holons, and
$-1/2$ spinons. A property of crucial importance for the pseudofermion
theory is that for the bare-momentum distribution deviations
corresponding to the PS excited energy eigenstates the energy terms
$E_i$ on the right-hand side of Eq. (\ref{EF}) associated
with the excitations generated by the elementary processes (A)
and (B) are of order $[1/L]^{i-1}$. Therefore, for the
pseudofermion-theory thermodynamic limit only the $i=1$ energy
term (\ref{E1}) is finite, all $i>1$ energy terms vanishing as
$[1/L]^{i-1}$. But according to the above considerations, since
the energy contributions in the bare-momentum
distribution-function deviations of order equal or larger than two
vanish in the thermodynamic limit, the same occurs for the
corresponding contributions in the canonical-momentum
distribution-function deviations. Therefore, the fact
that for the PS occupancy configurations generated by the
elementary processes (A) and (B) all energy terms $E_i$ on the right-hand
side of Eq. (\ref{EF}) such that $i>1$ vanish as $[1/L]^{i-1}$ implies
that for the thermodynamic-limit pseudofermion theory only the
$i=1$ leading energy-deviation term is finite in that limit. This
is consistent with the non-interacting character of
the PS general energy spectrum of Eqs. (\ref{ESS}) for the
pseudofermions. Indeed, such a non-interacting character implies
that the energy-deviation spectrum is linear in the pseudofermion
canonical-momentum distribution-function deviations and given
by Eq. (\ref{E1pf}), as discussed above.

On the other hand, the above $f$ functions are associated with the
two-pseudoparticle residual interactions and have the same role as
those of Fermi-liquid theory. Indeed, for small values of the
energy and electronic densities $n$ and spin densities $m$ such
that $0<na<1$ and $0<ma<na$, respectively, the low-energy physics
is controlled by the residual two-pseudoparticle interactions
described by the $i=2$ terms (\ref{E2}) of the energy functional
(\ref{EF}). Importantly, in that limit the leading-order terms in
$1/L$ of that energy spectrum are of first order in $1/L$ and are
contained in both the energy terms (\ref{E1}) and (\ref{E2}). On
the other hand, for the pseudofermions the energy spectrum
(\ref{E1pf}) vanishes in such a limit, and the same information is
stored in the pseudofermion canonical-momentum (\ref{barqan})
through the functional $Q^{\Phi}_{\alpha\nu}(q)/2$ of Eq. (\ref{qcan1j}),
as further discussed below and in Ref. \cite{V}. [For the
pseudoparticles the functional $Q^{\Phi}_{\alpha\nu}(q)/2$
appears instead in the residual-interaction terms (\ref{E2}).]

In the low-energy Hilbert subspace
only the $c0$ and $s1$ pseudoparticle branches have finite
occupancies \cite{Carmelo91,Carmelo92}. In this case, as the limit
of vanishing density of pseudoparticles contributing to the
bare-momentum distribution-function deviations is approached, the
general energy functional (\ref{EF}) acquires the form of the
energy spectrum of a two-component $c\equiv c0$ and $s\equiv s1$
conformal field theory \cite{Frahm}. The conformal dimensions that
control the asymptotic of the low-energy correlations functions
are extracted from the finite-size energy corrections
\cite{Frahm}. These energy corrections can be obtained by the use
in the pseudoparticle energy terms (\ref{E1}) and (\ref{E2}) of
deviations $\Delta N_{c0} (q)$ and $\Delta N_{s1} (q)$ descriptive
of general low-energy excitations \cite{Carmelo91,Carmelo92}. In
spite that for the pseudofermion theory the energy-deviation
spectrum (\ref{E1pf}) has no residual-interaction terms and thus is
linear in the canonical-momentum distribution-function deviations, it
is found in Ref. \cite{CFT} that the above conformal dimensions arise
naturally from the pseudofermion momentum for low-energy excitations.

For the pseudofermion theory, the energy (\ref{om0}) on the
right-hand side of Eq. (\ref{E1pf}) controls the finite-energy
physics. The remaining energy terms refer to gapless
contributions, provided that the involved pseudofermions
correspond to the canonical-momentum values in the vicinity of the
energy-band arguments of Eq. (\ref{eplev0}). For most excited energy
eigenstates, the latter terms also lead to finite-energy contributions.
The typical value of the latter energy contributions is of the order of
the pseudofermion energy dispersion band-width per pseudofermion
involved in the excited energy eigenstates. Note that the
PS energy spectrum (\ref{E1pf}) can have any finite value
associated with the regions of the ($k,\,\omega$)-plane where the
finite-number-electron spectral functions have finite spectral weight
\cite{V,spectral}.

Provided that one considers only the contributions of first-order in the
canonical-momentum distribution function deviations, the
momentum deviation spectrum can be written in terms of the pseudofermion
canonical-momentum distribution function deviations. It is
given by,

\begin{equation}
\Delta P = {\pi\over a}\,\Delta M_{c,\,-1/2} +
\sum_{j=1}^{N_a}\,\Delta {\cal{N}}_{c0} ({\bar{q}}_j)
\,{\bar{q}}_j + \sum_{\nu
=1}^{\infty}\,\sum_{j=1}^{N^*_{s\nu}}\,\Delta {\cal{N}}_{s\nu}
({\bar{q}}_j) \,{\bar{q}}_j + \sum_{\nu =1}^{\infty}\,
\sum_{j=1}^{N^*_{c\nu}}\,\Delta {\cal{N}}_{c\nu}
({\bar{q}}_j)\,[{\pi\over a} -{\bar{q}}_j]  \, . \label{noPpf}
\end{equation}

When acting onto the PS, the ground-state normal-ordered 1D
Hubbard model and momentum operator can be written in terms of
pseudofermion, $-1/2$ Yang holon, and $-1/2$ HL spinon operators
as follows,

\begin{equation}
:\hat{H}: = \sum_{\alpha\nu}\,\sum_{j=1}^{N^*_{\alpha\nu}}
\,\epsilon_{\alpha\nu}({\bar{q}}_j)\,:f^{\dag
}_{{\bar{q}}_j,\,\alpha\nu}:\,:f_{{\bar{q}}_j,\,\alpha\nu}: +
\sum_{\alpha =c,s}\,\epsilon_{L\alpha,\,-1/2}\,{\hat{L}}_{\alpha
,\,-1/2}
 \, , \label{Hno1pf}
\end{equation}
and

\begin{eqnarray}
:\hat{P}:\, & = & \sum_{j=1}^{N_a}\,{\bar{q}}_j\,
:f^{\dag}_{{\bar{q}}_j,\,c0}:\,:f_{{\bar{q}}_j,\,c0}: + \sum_{\nu
=1}^{\infty}\,\sum_{j=1}^{N^*_{s\nu}}\,{\bar{q}}_j :f^{\dag
}_{{\bar{q}}_j,\,s\nu}:\,:f_{{\bar{q}}_j,\,s\nu}: \nonumber \\ & +
& \sum_{\nu
=1}^{\infty}\,\sum_{j=1}^{N^*_{c\nu}}\,[(1+\nu){\pi\over a}
-{\bar{q}}_j]\,:f^{\dag
}_{{\bar{q}}_j,\,c\nu}:\,:f_{{\bar{q}}_j,\,c\nu}: + {\pi\over
a}\,{\hat{L}}_{c ,\,-1/2} \, , \label{noPoppf}
\end{eqnarray}
respectively, where $N^*_{c0}=N_a$ and the operator
${\hat{L}}_{\alpha ,\,-1/2}$ is given in Eq. (39) of Ref.
\cite{I}. On the right-hand side of Eq. (\ref{Hno1pf}), the
pseudofermion energy bands are defined by Eqs. (C.15)-(C.21) of Ref.
\cite{I} and the $-1/2$ Yang holon and $-1/2$ HL spinon
energies read \cite{II} $\epsilon_{Lc,\,-1/2} = 2\mu$ and
$\epsilon_{Ls,\,-1/2} = 2\mu_0\,H$, respectively.

The ground-state normal-ordered Hamiltonian (\ref{Hno1pf}) and
momentum operator (\ref{noPoppf}) correspond to the energy and
momentum deviation spectra given in Eqs. (\ref{E1pf}) and (\ref{noPpf}),
respectively.

\subsection{WAVE-FUNCTION FACTORIZATION OF THE NORMAL-ORDERED 1D HUBBARD
MODEL}

It is well known that both the ground state wave function and the
wave function of the excited energy eigenstates of the 1D Hubbard model can in
the $U/t\rightarrow\infty$ limit be constructed as a product of a
spin-less fermion wave function and a squeezed spin wave function
\cite{Ogata,Penc95,Penc96}. In our pseudofermion language this
factorization means that in such a limit the expression of the
momentum and energy spectra of these states is linear in the
canonical-momentum distribution functions. It is
straightforward to show that for finite values of $U/t$ the
general energy spectrum defined by Eq. (\ref{E}) and Eq. (20) of
Ref. \cite{II} is not linear in such functions. Therefore, the
above type of factorization does not occur in general for the 1D
Hubbard model.

Fortunately, the evaluation of finite-number-electron spectral
functions can be achieved without the full factorization of the wave
functions. That problem can be solved by use of the ground-state
normal-ordered 1D Hubbard model. When expressed in terms of
pseudofermion operators, that Hamiltonian and associated momentum
operator are given in Eqs. (\ref{Hno1pf}) and (\ref{noPoppf}),
respectively. In the thermodynamic limit, there is a wave-function
factorization for the PS excited energy eigenstates. This factorization
follows from the linear dependence on the canonical-momentum
distribution-function deviations of the expressions (\ref{E1pf})
and (\ref{noPpf}) for the energy and momentum, respectively.
Therefore, the wave function of the energy eigenstates of the
normal-ordered Hamiltonian can be expressed in the PS as a product
of wave functions. Each wave function corresponds to a different
pseudofermion branch. In excited energy eigenstates with finite independent
$-1/2$ holon and independent $-1/2$ spinon occupancy, there is also a wave function
for these objects. In contrast, for the pseudoparticle
representation the energy functional (\ref{EF})-(\ref{E2})
includes bare-momentum distribution function deviation non-linear
terms associated with the pseudoparticle residual interactions.
For the pseudoparticle representation we cannot ignore such energy
terms because they control the low-energy physics
\cite{Carmelo91,Carmelo92}. The occurrence of these energy terms
mixes contributions from different branches. It follows that the
pseudoparticle energy spectrum is not additive in the $\alpha\nu$
pseudoparticle branch contributions, in contrast to the
pseudofermion energy spectrum given in Eqs. (\ref{E1pf}). Thus, in
this case the wave function of the excited energy eigenstates does not
factorize in the form of a product of pseudoparticle wave
functions.

The number of wave functions contributing to the factorized wave
function of a given energy eigenstate depends on the occupancy
configurations of that state. Only the $\alpha\nu$ pseudofermion
branches with finite occupancy in the state contribute to the wave
function. This contribution is in the form of a $\alpha\nu$ wave
function factor, as further discussed in Refs. \cite{V,V-1}. The same
applies to the occupancy of independent $-1/2$ holons and
independent $-1/2$ spinons.

\section{CONCLUDING REMARKS}

In this paper we introduced a pseudofermion operational
description for the ground-state normal-ordered 1D Hubbard model.
We found that in the thermodynamic limit the wave function of
excited energy eigenstates belonging to the PS where the finite-number-electron
excitations are contained factorizes for all values of $U/t$. This
factorization results from the absence of residual-interaction energy terms
for the pseudofermions whose occupancy configurations describe these
excited energy eigenstates. Our study included the introduction of the
pseudoparticle - pseudofermion unitary transformation and of an
operator algebra for both the pseudoparticles and pseudofermions.
In the PS the functional $Q^{\Phi}_{\alpha,\,\nu}(q)/L$
associated with that transformation exactly cancels the residual
interactions of the $\alpha\nu$ pseudoparticles, through a mechanism
similar to the usual Peierls substitution. The information recorded in
the pseudoparticle interactions is contained in that functional and
is transferred over to the
pseudofermion canonical momentum (\ref{barqan}). Moreover, we introduced
creation and annihilation operators for both the pseudoparticles
and pseudofermions and derived the anticommutation relations of
these operators. The pseudofermion anticommutation relations
play a central role in the study of the spectral and dynamical
properties \cite{V,V-1,spectral,super}.

The pseudofermion operator algebra introduced here is used in
Refs. \cite{V,V-1} in the construction of a pseudofermion dynamical
theory. That theory allows the evaluation of finite-number-electron
spectral-function expressions for all energy values. Furthermore,
the pseudofermion operational description is useful for the
further understanding of the exotic properties displayed by
low-dimensional materials. A preliminary application of the
pseudofermion dynamical theory to the study of the one-electron
spectral functions is presented in Refs.
\cite{spectral0,spectral}. The theoretical predictions of these
references describe quantitatively for the whole finite-energy
band width the one-electron removal spectral lines observed by
photoemission experiments in the quasi-1D organic compound
TTF-TCNQ. Also the predictions of Ref. \cite{super} are
consistent with the phase diagram observed in a series of organic
compounds. Recently, the one-electron problem investigated in
Refs. \cite{spectral0,spectral} was studied by the dynamical
density matrix renormalization group method in Ref.
\cite{Eric}. The studies of the latter reference reached results
similar to those of the former references. The pseudofermion
description introduced here is also of interest for the understanding
the spectral properties of the new quantum systems
described by cold fermionic atoms on an optical lattice.

\begin{acknowledgments}
I dedicate this work to the memory of Alexander A. Ovchinnikov, with whom
I had illuminating preliminary discussions relevant for the results obtained here. I
thank Karlo Penc for many useful and stimulating discussions concerning the
issues studied in this paper. I also thank Daniel Bozi, Ant\^onio Castro
Neto, Francisco (Paco) Guinea, Patrick A. Lee, Lu\'{\i}s
Miguel Martelo, and Pedro Sacramento, for illuminating discussions. I am
grateful for the hospitality and support of MIT and the financial support
of the Gulbenkian Foundation and Fulbright Commission.
\end{acknowledgments}
\appendix

\section{THE RAPIDITY TWO-PSEUDOFERMION PHASE SHIFTS $\bar{\Phi}_{\alpha\nu,\,\alpha'\nu'}
\left(r, r'\right)$}

Here we provide the set of integral equations which define the
rapidity two-pseudofermion phase shifts $\bar{\Phi
}_{\alpha\nu,\,\alpha'\nu'}\left(r,r'\right)$ in
units of $\pi$ on the right-hand
side of Eqs. (\ref{Phi-barPhi}). Let us start by introducing the
following {\it Fermi surface} parameters $r^0_c = 4t\,\sin Q/U$
and $r^0_s = 4t\,B/U$ where the parameters $Q$ and $B$ are defined
in Ref. \cite{V-1}. In order to derive
the integral equations which define the rapidity two-pseudofermion
phase shifts $\bar{\Phi
}_{\alpha\nu,\,\alpha'\nu'}\left(r,r'\right)$, we first use in
Eqs. (13)-(16) of Ref. \cite{I} the bare-momentum distribution
functions of the general form (\ref{N0DNq}). Expansion of the
obtained equations up to first order in the bare-momentum
distribution function deviations on the right-hand side of Eqs.
(\ref{DNq}) and (\ref{Qcan1j}) leads to expression
(\ref{Phi-barPhi}) with the two-pseudofermion phase shift
$\bar{\Phi }_{\alpha\nu,\,\alpha'\nu'} (r ,\,r')$ uniquely defined
by the integral equations given below. A first group of
two-pseudofermion phase shifts obey integral equations by their
own. These equations read,

\begin{equation}
\bar{\Phi }_{s1,\,c0}\left(r,r'\right) = -{1\over{\pi}}{\rm
arc}{\rm tan}(r-r') + \int_{-r^0_s}^{r^0_s}
dr''\,G(r,r'')\,{\bar{\Phi }}_{s1,\,c0}\left(r'',r'\right) \, ,
\label{Phis1c}
\end{equation}

\begin{equation}
\bar{\Phi }_{s1,\,c\nu}\left(r,r'\right) =  -
{1\over{\pi^2}}\int_{-r^0_c}^{r^0_c} dr''{{\rm arc}{\rm tan}
\Bigl({r''-r'\over\nu}\Bigr)\over{1+(r-r'')^2}} +
\int_{-r^0_s}^{r^0_s} dr''\,G(r,r'')\,{\bar{\Phi
}}_{s1,\,c\nu}\left(r'',r'\right) \, , \label{Phis1cn}
\end{equation}
and

\begin{eqnarray}
\bar{\Phi }_{s1,\,s\nu}\left(r,r'\right) & = & {\delta_{1
,\,\nu}\over\pi}\,\arctan\Bigl({r-r'\over 2}\Bigl) + {(1-\delta_{1
,\,\nu})\over\pi}\Bigl\{ \arctan\Bigl({r-r'\over \nu-1}\Bigl) +
\arctan\Bigl({r-r'\over \nu+1}\Bigl)\Bigr\} \nonumber \\ & - &
{1\over{\pi^2}}\int_{-r^0_c}^{r^0_c} dr''{{\rm arc}{\rm tan}
\Bigl({r''-r'\over\nu}\Bigr)\over{1+(r-r'')^2}} +
\int_{-r^0_s}^{r^0_s} dr''\,G(r,r'')\,{\bar{\Phi
}}_{s1,\,s1}\left(r'',r'\right) \, . \label{Phis1sn}
\end{eqnarray}
Here the kernel $G(r,r')$ is given by \cite{Carmelo92},

\begin{equation}
G(r,r') = - {1\over{2\pi}}\left[{1\over{1+((r-r')/2)^2}}\right]
\left[1 - {1\over 2}
\left(t(r)+t(r')+{{l(r)-l(r')}\over{r-r'}}\right)\right] \, ,
\label{G}
\end{equation}
where $t(r) = {1\over{\pi}}\left[{\rm arc}{\rm tan}(r + r^0_c) -
{\rm arc}{\rm tan}(r -r^0_c)\right]$ and $l(r) =
{1\over{\pi}}\left[ \ln (1+(r + r^0_c)^2) - \ln (1+(r
-r^0_c)^2)\right]$. A second group of two-pseudofermion phase
shifts are expressed in terms of the basic functions given in Eqs.
(\ref{Phis1c})-(\ref{Phis1sn}) as follows,

\begin{equation}
\bar{\Phi }_{c0,\,c0}\left(r,r'\right) =
{1\over{\pi}}\int_{-r^0_s}^{r^0_s} dr''{\bar{\Phi
}_{s1,\,c0}\left(r'',r'\right) \over {1+(r-r'')^2}} \, ,
\label{Phicc}
\end{equation}

\begin{equation}
\bar{\Phi }_{c0,\,c\nu}\left(r,r'\right) = -{1\over{\pi}}{\rm
arc}{\rm tan}\Bigl({r-r'\over \nu}\Bigr) +
{1\over{\pi}}\int_{-r^0_s}^{r^0_s} dr''{\bar{\Phi
}_{s1,\,c\nu}\left(r'',r'\right) \over {1+(r-r'')^2}} \, ,
\label{Phiccn}
\end{equation}
and

\begin{equation}
\bar{\Phi }_{c0,\,s\nu}\left(r,r'\right) = -{1\over{\pi}}{\rm
arc}{\rm tan}\Bigl({r-r'\over \nu}\Bigr) + {1\over{\pi}}
\int_{-r^0_s}^{r^0_s} dr''{\bar{\Phi
}_{s1,\,s\nu}\left(r'',r'\right) \over {1+(r-r'')^2}} \, .
\label{Phicsn}
\end{equation}

Finally, the remaining two-pseudofermion phase shifts can be
expressed either in terms of the functions
(\ref{Phicc})-(\ref{Phicsn}) only,

\begin{equation}
{\bar{\Phi }}_{c\nu,\,c0}\left(r,r'\right) = {1\over{\pi}}{\rm
arc}{\rm tan}\Bigl({r-r'\over {\nu}}\Bigr) -
{1\over{\pi}}\int_{-r^0_c}^{r^0_c} dr''{{\bar{\Phi
}}_{c0,\,c0}\left(r'',r'\right) \over {\nu[1+({r-r''\over
{\nu}})^2]}} \, , \label{Phicnc}
\end{equation}

\begin{equation}
\bar{\Phi }_{c\nu,\,c\nu'}\left(r,r'\right) =
{1\over{2\pi}}\Theta_{\nu,\,\nu'}(r-r') -
{1\over{\pi}}\int_{-r^0_c}^{r^0_c} dr''{\bar{\Phi
}_{c0,\,c\nu'}\left(r'',r'\right) \over
{\nu[1+({r-r''\over\nu})^2]}} \, , \label{Phicncn}
\end{equation}
and

\begin{equation}
\bar{\Phi }_{c\nu,\,s\nu'}\left(r,r'\right) = -
{1\over{\pi}}\int_{-r^0_c}^{r^0_c} dr''{\bar{\Phi
}_{c0,\,s\nu'}\left(r'',r'\right) \over
{\nu[1+({r-r''\over\nu})^2]}} \, , \label{Phicnsn}
\end{equation}
or both in terms of the basic functions
(\ref{Phis1c})-(\ref{Phis1sn}) and of the phase shifts
(\ref{Phicc})-(\ref{Phicsn}),

\begin{equation}
{\bar{\Phi }}_{s\nu ,\,c0}\left(r,r'\right) = - {{\rm arc}{\rm
tan}\Bigl({r-r'\over {\nu}}\Bigr)\over \pi} +
{1\over{\pi}}\int_{-r^0_c}^{r^0_c} dr''{{\bar{\Phi
}}_{c0,\,c0}\left(r'',r'\right) \over {\nu[1+({r-r''\over
\nu})^2]}} - \int_{-r^0_s}^{r^0_s} dr''{\bar{\Phi
}}_{s1,\,c0}\left(r'',r'\right)
{\Theta^{[1]}_{\nu,\,1}(r-r'')\over{2\pi}} \, ; \hspace{0.5cm} \nu
> 1 \, , \label{Phisnc}
\end{equation}

\begin{equation}
{\bar{\Phi }}_{s\nu ,\,c\nu'}\left(r,r'\right) =
{1\over{\pi}}\int_{-r^0_c}^{r^0_c} dr''{{\bar{\Phi
}}_{c0,\,c\nu'}\left(r'',r'\right) \over {\nu[1+({r-r''\over
\nu})^2]}} - \int_{-r^0_s}^{r^0_s} dr''{\bar{\Phi
}}_{s1,\,c\nu'}\left(r'',r'\right)
{\Theta^{[1]}_{\nu,\,1}(r-r'')\over {2\pi}} \, ; \hspace{0.5cm}
\nu > 1 \, , \label{Phisncn}
\end{equation}
and

\begin{equation}
{\bar{\Phi }}_{s\nu ,\,s\nu'}\left(r,r'\right) =
{\Theta_{\nu,\,\nu'}(r-r')\over{2\pi}} +
{1\over{\pi}}\int_{-r^0_c}^{r^0_c} dr''{{\bar{\Phi
}}_{c0,\,s\nu'}\left(r'',r'\right) \over {\nu[1+({r-r''\over
\nu})^2]}} - \int_{-r^0_s}^{r^0_s} dr''{\bar{\Phi
}}_{s1,\,s\nu'}\left(r'',r'\right)
{\Theta^{[1]}_{\nu,\,1}(r-r'')\over{2\pi}} \, ; \hspace{0.5cm} \nu
> 1 \, . \label{Phisnsn}
\end{equation}
In the above two-pseudofermion phase shift expressions the
functions $\Theta_{\nu,\,\nu'}(x)$ and
$\Theta^{[1]}_{\nu,\,\nu'}(x)$ are given in Eqs. (B5) and (C22) of
Ref. \cite{I}, respectively. In spite of the different notation
and except for simplifications introduced here as a result of some
integrations performed analytically, the integral equations
(\ref{Phis1c})-(\ref{Phisnsn}) are equivalent to the system of
coupled integral equations (B30)-(B40) of Ref. \cite{Carmelo97}.

\section{THE $\alpha\nu$ PSEUDOPARTICLE - PSEUDOFERMION UNITARY OPERATOR}

Here we confirm that in the PS the $\alpha\nu$ pseudoparticle - pseudofermion
operator ${\hat{V}}_{\alpha\nu}$ that obeys Eq. (\ref{f}) is
unitary and given by expression (\ref{Van}). While that expression refers
to the PS, for any energy eigenstate of the full Hilbert space
there is a one-to-one correspondence between the set of specific
discrete values of the rapidity momentum $\{k_j\} = \{k(q_j)\}$
such that $j=1,...,N_a$ and the set of $c0$ band discrete
bare-momentum values $\{q_j\}$. Moreover, there is a one-to-one
correspondence between the set of specific discrete values of
each $\alpha\nu$-branch rapidity $\{\Lambda_{j,\,\alpha\nu}\}=
\{\Lambda_{\alpha\nu}(q_j)\}$ such that $j=1,...,N^*_{\alpha\nu}$
and the set of $\alpha\nu$ band discrete bare-momentum values $\{q_j\}$.
These branches are such that $\alpha=c,\,s$ and $\nu =1,2,3,...$.
This correspondence is fully defined by the integral equations (13)-(16)
of Ref. \cite{I}, which refer to a functional representation of
the thermodynamic Bethe-ansatz equations introduced by Takahashi
\cite{Takahashi}. The rapidity-momentum functional is real and the rapidity
functionals are the real part of Takahashi's ideal strings \cite{Takahashi,I}.

Two alternative equivalent and complete descriptions for each energy
eigenstate correspond to: (a) occupancy configurations of the above
sets of discrete bare-momentum values $\{q_j\}$ for the $\alpha\nu$
bands with finite bare-momentum occupancy for the state plus occupancies of the
$L_{c,\,-1/2}$ and $L_{s,\,-1/2}$ numbers; (b) occupancy configurations
of the above set of discrete numbers $\{k_j\}$ and
$\{\Lambda_{j,\,\alpha\nu}\}$ for the $\alpha\nu$ branches with finite
occupancy of these numbers for the state plus occupancies of the
$L_{c,\,-1/2}$ and $L_{s,\,-1/2}$ numbers. Since both these descriptions
describe the same energy eigenstates, there is a uniquely defined
transformation connecting the two alternative representations. That
transformation refers to the whole Hilbert space.

For the PS, that correspondence assumes the simple form given in Eq.
(\ref{FL}). There $k^0(q_j)$ and $\Lambda^0_{\alpha\nu}(q_j)$ are the
initial ground-state rapidity-momentum and rapidity functions,
respectively, whose inverse functions are given in
Ref. \cite{V-1} and ${\bar{q}} (q)$ is the canonical-momentum
provided in Eq. (\ref{barqan}). It follows from Eq. (\ref{FL}) that
for the PS the $\alpha\nu$ pseudoparticle - $\alpha\nu$ pseudofermion
transformation described by Eqs. (\ref{pfermions}) and (\ref{f}) fully controls the
above bare-momentum - rapidity-momentum/rapidity transformation.
Since the $\alpha\nu$ pseudoparticle - pseudofermion transformation
connects two alternative representations for the complete set of
orthogonality and normalized energy eigenstates that span the PS, the
operator ${\hat{V}}_{\alpha\nu}$ of Eq. (\ref{f}) must be unitary in
the PS, as confirmed below.

Equation (\ref{Van}) leads to,

\begin{equation}
{\hat{V}}^{\dag}_{\alpha\nu} =
\exp\bigl\{\sum_{q_j}\,[\,b_{q_j+\delta (q_j),\,\alpha\nu}^{\dag}
- b_{q_j,\,\alpha\nu}^{\dag}]\,b_{q_j,\,\alpha\nu}\Bigr\} \, ; \hspace{1cm}
\delta (q_j) =Q^{\Phi}_{\alpha\nu}(q_j)/L \, .
\label{Van+}
\end{equation}
For the PS, the functional $\delta (q_j) =Q^{\Phi}_{\alpha\nu}(q_j)/L$
given in Eq. (\ref{qcan1j}) is of the order $1/L$. Thus, within the
thermodynamic limit that the pseudofermion description corresponds
to, we use the following representation for the operator
$[\,b_{q_j+\delta (q_j),\,\alpha\nu}^{\dag}
- b_{q_j,\,\alpha\nu}^{\dag}]$ in terms of the continuum
bare-momentum $q$,

\begin{equation}
[\,b_{q_j+\delta (q_j),\,\alpha\nu}^{\dag}
- b_{q_j,\,\alpha\nu}^{\dag}] = \delta (q)
\,{\partial\over\partial q}\,b_{q,\,\alpha\nu}^{\dag} \, .
\label{b-b-db}
\end{equation}
Next, by expanding the exponential of Eq. (\ref{Van})
we find that to first-order in $1/L$ the operator (\ref{Van+}) obeys
the following equation,

\begin{equation}
{\hat{V}}^{\dag}_{\alpha\nu}\,b^{\dag
}_{q_j,\,\alpha\nu} = b^{\dag
}_{q_j+\delta (q_j),\,\alpha\nu}\,{\hat{V}}^{\dag}_{\alpha\nu}
\, . \label{Vb}
\end{equation}
This equation is equivalent to Eq. (\ref{f}). Therefore, this confirms
that for the PS the operators (\ref{Van}) and (\ref{Van+}) indeed
obey the relation (\ref{f}). Let us next show that in the PS
these operators are unitary.

The operator (\ref{Van+}) can be rewritten as,

\begin{equation}
{\hat{V}}^{\dag}_{\alpha\nu} = e^{i{\hat{G}}_{\alpha\nu}} \, ;
\hspace{1cm} {\hat{G}}_{\alpha\nu} = -i
\sum_{q_j}\,[\,b_{q_j+\delta (q_j),\,\alpha\nu}^{\dag}
- b_{q_j,\,\alpha\nu}^{\dag}]\,b_{q_j,\,\alpha\nu} \, .
\label{Van+G}
\end{equation}
If the operator ${\hat{G}}_{\alpha\nu}$ is hermitian then the operator
${\hat{V}}_{\alpha\nu}$ is unitary. Let us show that the invariance of
the $\alpha\nu$ pseudoparticle number operator (\ref{Nop})
under the transformation (\ref{pfermions}), which implies that it equals the
$\alpha\nu$ pseudofermion number operator (\ref{Nbarop}), also implies the
hermitian character of the operator ${\hat{G}}_{\alpha\nu}$ of
Eq. (\ref{Van+G}).

Taking the transpose of the operator ${\hat{G}}_{\alpha\nu}$ of
Eq. (\ref{Van+G}) leads to,

\begin{equation}
{\hat{G}}^{\dag}_{\alpha\nu} = i
\sum_{q_j}\,b_{q_j,\,\alpha\nu}^{\dag}[\,b_{q_j+\delta (q_j),\,\alpha\nu}
- b_{q_j,\,\alpha\nu}] \, .
\label{G-}
\end{equation}
By using the transpose of Eq. (\ref{b-b-db}) we rewrite the operator
(\ref{G-}) as,

\begin{eqnarray}
{\hat{G}}^{\dag}_{\alpha\nu} & = & i {L\over 2\pi}
\int dq\,\delta (q)\,b_{q,\,\alpha\nu}^{\dag}\,{\partial
\over\partial q}\,b_{q,\,\alpha\nu} = -i
{L\over 2\pi}
\int dq\,{\partial\over\partial q}\Bigl[\,\delta (q)\,
b_{q,\,\alpha\nu}^{\dag}\Bigr]\,b_{q,\,\alpha\nu}
+ i {L\over 2\pi}
\int dq\,{\partial \over\partial q}\Bigl[\,\delta (q)\,
b_{q,\,\alpha\nu}^{\dag}\,b_{q,\,\alpha\nu}\Bigr]\nonumber \\
& = &
{\hat{G}}_{\alpha\nu} -i
{L\over 2\pi}
\int dq\,\Bigl[{\partial\over\partial q}\,\delta (q)\Bigr]\,
b_{q,\,\alpha\nu}^{\dag}\,b_{q,\,\alpha\nu}
+ i {L\over 2\pi}
\int dq\,{\partial \over\partial q}\Bigl[\,\delta (q)\,
b_{q,\,\alpha\nu}^{\dag}\,b_{q,\,\alpha\nu}\Bigr] \, .
\label{G-cont}
\end{eqnarray}
where

\begin{equation}
{\hat{G}}_{\alpha\nu} = -i
{L\over 2\pi}
\int dq\,\delta (q)\,{\partial
\over\partial q}\Bigl[\,b_{q,\,\alpha\nu}^{\dag}\Bigr]\,b_{q,\,\alpha\nu} = -i
\sum_{q_j}\,[\,b_{q_j+\delta (q_j),\,\alpha\nu}^{\dag}
- b_{q_j,\,\alpha\nu}^{\dag}]\,b_{q_j,\,\alpha\nu} \, .
\label{Van+G-cont}
\end{equation}
In equations (\ref{G-cont}) and (\ref{G-cont}) we have replaced the bare-momentum
summations by integrals over the whole $q$ domain of the corresponding
$\alpha\nu$ band.

We emphasize that the operator ${\hat{G}}_{\alpha\nu}$ of
Eq. (\ref{Van+G}) is hermitian provided that the two last
terms on the right-hand side of the last line of Eq. (\ref{G-cont})
vanish. These terms can be rewritten as,

\begin{eqnarray}
& & -i {L\over 2\pi}
\int dq\,\Bigl[{\partial\over\partial q}\,\delta (q)\Bigr]\,
b_{q,\,\alpha\nu}^{\dag}\,b_{q,\,\alpha\nu}
+ i {L\over 2\pi}
\int dq\,{\partial \over\partial q}\Bigl[\,\delta (q)\,
b_{q,\,\alpha\nu}^{\dag}\,b_{q,\,\alpha\nu}\Bigr] \nonumber \\
& = & i {L\over 2\pi}
\int dq\,\delta (q)\,{\partial\over\partial q}\Bigl[\,
b_{q,\,\alpha\nu}^{\dag}\,b_{q,\,\alpha\nu}\Bigr]
- i {L\over 2\pi}
\int dq\,{\partial \over\partial q}\Bigl[\,\delta (q)\,
b_{q,\,\alpha\nu}^{\dag}\,b_{q,\,\alpha\nu}\Bigr]
+ i {L\over 2\pi}
\int dq\,{\partial \over\partial q}\Bigl[\,\delta (q)\,
b_{q,\,\alpha\nu}^{\dag}\,b_{q,\,\alpha\nu}\Bigr] \nonumber \\
& = & i \sum_{q_j}\Bigl[\,\hat{N}_{\alpha\nu}(q_j+\delta (q_j))
-\hat{N}_{\alpha\nu}(q_j)\Bigr] = i \Bigl[\,\sum_{{\bar{q}}_j}\hat{\cal{N}}_{\alpha\nu}({\bar{q}}_j)
-\sum_{q_j}\hat{N}_{\alpha\nu}(q_j)\Bigr] =
i\Bigl[\,\hat{\cal{N}}_{\alpha\nu}
-\hat{N}_{\alpha\nu}\Bigr] = 0 \, ,
\label{0-G}
\end{eqnarray}
where here the $\alpha\nu$ pseudofermion number operator (\ref{Nbarop})
was called $\hat{\cal{N}}_{\alpha\nu}$ and $\hat{N}_{\alpha\nu}$
is the $\alpha\nu$ pseudoparticle number operator (\ref{Nop}).
Since these operators are invariant under the transformation (\ref{pfermions})
they are the same operator, what justifies the vanishing of the
operator terms (\ref{0-G}). This shows that the operator ${\hat{G}}_{\alpha\nu}$ of
Eq. (\ref{Van+G}) is hermitian and thus that the operator ${\hat{V}}_{\alpha\nu}$
(\ref{Van}) is unitary.


\end{document}